\documentclass[journal, a4paper]{IEEEtran}

\usepackage{amsmath,amssymb,amsfonts,amsthm,mathtools}

\usepackage[utf8]{inputenc}
\usepackage[T1]{fontenc}
\usepackage{hyperref}
\usepackage{url}
\usepackage{booktabs}
\usepackage{nicefrac}
\usepackage{microtype}
\usepackage{xcolor}
\usepackage{graphicx}
\usepackage{subfigure}
\usepackage{multirow}
\usepackage{cite}
\usepackage{bm}
\usepackage{cases}
\usepackage{cuted}\stripsep-3pt
\usepackage{algorithm}
\usepackage{algorithmic}
\usepackage{color}
\usepackage[normalem]{ulem}
\usepackage{float}
\usepackage{stfloats}

\newcommand{\kap}{\kappa}          
\newcommand{\alp}{\alpha}          

\newcommand{\thetadiv}{\theta_{\mathrm{div}}}
\newcommand{\onaxis}{P(0)}         

\begin{document}
\title{Toroidal-OAM Hybrid Beam for Divergence-Suppressed OAM Wireless Communication}


\author{
\IEEEauthorblockN{Runyu Lyu, \textit{Member, IEEE}, Wenchi Cheng, \textit{Senior Member, IEEE}, and Xiang Fang, \textit{Student Member, IEEE}}

\thanks{Runyu Lyu and Xiang Fang are with the School of Electronics and Information Engineering, Nanjing University of Information Science and Technology, Nanjing, 210044, China (e-mail: runyulyu@nuist.edu.cn).

Wenchi Cheng is with the School of Telecommunications Engineering, Xidian University, Xi'an, 710071, China (e-mail: wccheng@xidian.edu.cn).
}
}

\IEEEtitleabstractindextext{%
\begin{abstract}
Orbital angular momentum (OAM) beams provide a spatially independent multiplexing dimension for wireless communications, but their donut-shaped intensity pattern suffers from divergence that increases with the OAM mode order. This divergence leads to significant power capture loss at receivers with a finite aperture and creates unequal link budgets across multiplexed OAM channels. We propose a toroidal-OAM hybrid beam that combines the angular confinement and strong longitudinal electric field (far-field features of toroidal electromagnetic pulses produced by a coaxial horn) with a conventional OAM beam. This hybrid beam simultaneously lowers the angular spread and delivers non-zero on-axis intensity. An analytical far-field model is built as a weighted superposition of the OAM uniform circular array (UCA) radiation, a super-Gaussian toroidal angular envelope, and a zeroth-order Bessel longitudinal field component. The model is controlled by three dimensionless parameters: toroidal confinement, coupling strength, and longitudinal coefficient, each linked to measurable hardware attributes. Numerical results at 3~GHz with a $5\lambda$ aperture show a 12--41\% reduction in divergence angle for OAM modes $\ell = 1$--$4$, with modes $\ell \geq 5$ transitioning to a fully center-peaked beam (its intensity peak on the axis). The suppression ratio remains unchanged when carrier frequency or aperture size varies (under geometric scaling that holds the dimensionless parameters $kR$, $\kappa$, $\alpha$, $\beta$, and $\ell$ fixed). Communication link evaluation reveals 6.4--30.9~dB gain in received signal-to-noise ratio, more than 108\% extension in the fixed-efficiency link-budget range equivalent, and a reduction of multi-mode power imbalance by up to five orders of magnitude. The central intensity null is filled to 8--100\% of the peak value, and the hybrid beam does not mix OAM modes, with the intended-mode power fraction at the ring staying above $0.54$ for modes where the ring is well-defined. An OAM mode-division multiplexing capacity analysis at $z=500\lambda$ demonstrates up to $2.1\times$ ($107\%$) capacity enhancement under the hybrid beam with each scheme normalized to its own peak. Full-wave electromagnetic simulation in ANSYS High Frequency Structure Simulator (HFSS) corroborates the beam model, confirming that the hybrid launcher preserves the OAM vortex phase structure while achieving measurable divergence reduction compared with a UCA-only baseline. These findings establish toroidal beam engineering as a practical tool for OAM-based wireless links.
\end{abstract}

}

\maketitle
\IEEEdisplaynontitleabstractindextext
\IEEEpeerreviewmaketitle


\section{Introduction}
\IEEEPARstart{O}{rbital} angular momentum (OAM) of electromagnetic waves introduces a new spatial degree of freedom for wireless communications, complementing conventional resources such as frequency, time, and polarization~\cite{allen1992orbital, thide2007utilization, tamburini2012encoding}. OAM originates from a helical phase front $\exp(i\ell\phi)$, where $\phi$ denotes the azimuth angle, that produces a donut-shaped intensity profile with a dark central null. Waves carrying different integer OAM mode orders $\ell$ are mutually orthogonal, which enables OAM mode-division multiplexing to multiply spectral efficiency relative to single-beam transmission over the same aperture, consistent with the LOS-MIMO equivalence of~\cite{edfors2012LoSMIMO}, without claiming a capacity bound beyond MIMO~\cite{yao2011orbital, cheng2019orbital, chen2021oam, trichili2019roadmap}. Recent demonstrations in the radio frequency domain, including multi-gigabit-per-second links and kilometer-scale outdoor experiments, confirm the practical potential of OAM multiplexing~\cite{tamburini2012encoding, yan2014high, zhao2017performance}.  This potential extends to optical free-space and space-borne links: OAM multiplexing has been reported at $100$~Gbps~\cite{dutta2021fso100} and $640$~Gbps~\cite{dutta2022fso640} in free-space optical transmission systems, and channel models for terahertz OAM-based inter-satellite communications have been established~\cite{gao2025channel}. Complementary system-level advances such as fractal and quasi-fractal uniform circular arrays (UCAs), cooperative transmission, and reconfigurable intelligent surface (RIS)-assisted OAM further strengthen the viability of OAM-based communications~\cite{jin2025ndquasifractal,cheng2024breaking, cheng2024cooperative, lyu2021oamnfc, cheng2024fractal, cheng2024quasifractal, lyu2026eit}.

 The LOS focus of this paper does not imply short distances: radio-frequency OAM links have been demonstrated outdoors over hundreds of meters, e.g., $442$~m at $2.414$~GHz in~\cite{tamburini2012encoding}, in addition to indoor millimeter-wave demonstrations~\cite{yan2014high} and array-level performance studies~\cite{zhao2017performance}. The achievable distance is limited by the mode-dependent divergence rather than by the LOS assumption itself, which is exactly what the proposed beam suppresses. OAM multiplexing is LOS-sensitive because the receive-aperture helical phase must be preserved; in NLOS-rich environments it can be enabled by RIS-assisted reflection~\cite{lyu2022joint}, with cooperative reception for aligned OAM links studied in~\cite{cheng2024cooperative}. The paper targets the LOS segment, and the RIS-assisted NLOS extension is left for future work.

Despite these advances, OAM beams face a fundamental physical limitation: their divergence angle is strongly mode-dependent. For a uniform circular array of radius $R$, the far-field angular power of an OAM mode $\ell$ is shaped by a Bessel function whose first maximum determines the divergence half-angle. Because the first peak of the Bessel function moves outward as $\ell$ grows, the angular spread of the beam widens sharply with increasing mode order. At a typical aperture of five wavelengths ($5\lambda$) and 3~GHz, the divergence angle spans from a few degrees to nearly eighteen degrees across the first eight modes, with a coefficient of variation around 47\%. This angular broadening results in severe power capture loss at finite-aperture receivers and creates unequal effective path losses across multiplexed OAM channels. Consequently, link budgets become unbalanced, higher-order modes are susceptible to misalignment, and only the lowest few modes can be used efficiently over practical distances.

The divergence angle is a first-order KPI of an OAM link: it determines finite-aperture power capture and hence per-mode received SNR, spectral efficiency, and bit error rate. Its mode-dependent loss beyond free-space path loss, rather than the LOS assumption itself, limits the multiplexable modes at a given distance~\cite{padgett2015divergence, cheng2019orbital}. The divergence half-angle locates the annular intensity maximum, whereas beamwidth measures that lobe's angular width; both follow the aperture-diffraction scale (Section~\ref{sec:analysis}).

Various methods have been proposed to control OAM divergence, including lens collimation, array phase optimization, perfect vortex beam (PVB) generation, and achromatic perfect vortex beams~\cite{tao2024novel, ostrovsky2013generation, zhao2025stable, xia2025ultrabroadband}. While these solutions offer some degree of divergence equalization, they generally rely on diffraction engineering within the conventional transverse electromagnetic framework and do not exploit the intrinsic robustness of nonseparable electromagnetic states.

Toroidal electromagnetic pulses, often called flying doughnuts, represent a different class of solutions to Maxwell's equations in which the electric and magnetic field lines form nested toroidal surfaces~\cite{hellwarth1996focused, raybould2016toroidal}. These pulses exhibit resilient propagation, a strong longitudinal field component, and skyrmionic field textures that resist diffractive spreading~\cite{zdagkas2022observation, wang2024observation, wang2024propagation}. The complete toroidal pulse topology (space-time nonseparability, skyrmionic textures) requires broadband pulsed excitation. Under the narrowband continuous-wave (CW) operation of a communication system, the coaxial horn's far-field radiation retains two features inherited from its toroidal-pulse-generating capability: angular confinement (a super-Gaussian angular envelope) and a strong longitudinal electric field component. These features can be understood as the monochromatic residue of the broadband toroidal pulse: the same antenna geometry and poloidal current paths that produce toroidal pulses under pulsed excitation yield angular confinement and longitudinal field under CW excitation. The recent demonstration of hybrid electromagnetic toroidal vortices~\cite{hetv2025hybrid} shows that toroidal topology and OAM can coexist in a single wavepacket, yet the potential of these toroidal-pulse-derived features for divergence control in wireless communication has not been explored.

OAM mode $\ell$ is an orbital-angular-momentum eigenstate carrying OAM proportional to $\ell$; the magnitude and sign of $\ell$ specify the number of helical surfaces and their handedness. Mode $\ell = 0$ has no helical phase and peaks on axis; the order-one modes are opposite-handed single helices, while increasing nonzero order enlarges the ring around the central phase singularity. Different azimuthal windings make the modes orthogonal parallel spatial channels. A UCA generates mode $\ell$ using equal-amplitude feeds whose phase advances $\ell$ full turns around the aperture. Reprogramming the digital phase-shifter network switches modes; superimposing these feed profiles multiplexes them, and conjugate spatial Fourier processing separates them at the receiver (Section~\ref{sec:capacity}).

In this paper, we propose a toroidal-OAM hybrid beam that harnesses these toroidal-pulse-derived features (angular confinement and longitudinal electric field) to suppress angular divergence and fill the central intensity null inherent to conventional OAM beams. The hybrid beam is constructed as a weighted superposition of a standard OAM-UCA field, an angularly filtered OAM component via a super-Gaussian toroidal envelope, and a longitudinal zeroth-order Bessel pedestal. We formulate a closed-form far-field model parameterized by three physically meaningful quantities: the toroidal confinement parameter, the hybrid coupling coefficient, and the longitudinal coupling strength. Each parameter is directly mapped to measurable hardware properties. Using this model, we evaluate divergence reduction, on-axis field filling, mode purity, communication link metrics, and OAM mode-division multiplexing capacity across OAM orders $\ell = 1$--$8$ at microwave frequencies. The results show 12--41\% divergence suppression for $\ell = 1$--$4$ with higher-order modes becoming fully center-peaked, 6.4--30.9~dB signal-to-noise ratio gain, and up to five orders of magnitude reduction in multi-mode power imbalance, while the hybrid beam does not mix OAM modes and the intended-mode power fraction at the ring stays above $0.54$ for modes where the OAM ring is well-defined. The suppression ratio is independent of carrier frequency and aperture size (under geometric scaling that holds the dimensionless parameters $kR$, $\kappa$, $\alpha$, $\beta$ fixed). Full-wave electromagnetic simulation in ANSYS High Frequency Structure Simulator (HFSS) corroborates the analytical model. These findings establish toroidal beam engineering as a practical paradigm for OAM-based wireless links.

 The hybrid is an RF-front-end modification reusing conventional UCA elements and phase shifters; $\ell = 0$ gives a broadside beam with standard DFT demultiplexing~\cite{edfors2012LoSMIMO}. Reprogramming per-element phase-shifter/vector-modulator feeds supporting superposed phase ramps switches between plane-wave beamforming and toroidal-OAM multiplexing.

The rest of the paper is organized as follows. Section~II describes the system model. Section~III presents the beam analysis, including the conventional OAM-UCA far-field pattern, the toroidal angular filter, the hybrid beam formulation, link budget, mode purity, and physical parameter extraction. Section~IV derives the OAM mode-division multiplexing channel capacity. Section~V provides comprehensive numerical results and discussion. Section~VI presents a full-wave HFSS simulation validation of the hybrid beam launcher. Finally, Section~VII concludes the paper.


\section{System Model}
\label{sec:system_model}



\begin{figure}[htbp]
\centering
\includegraphics[scale=0.55]{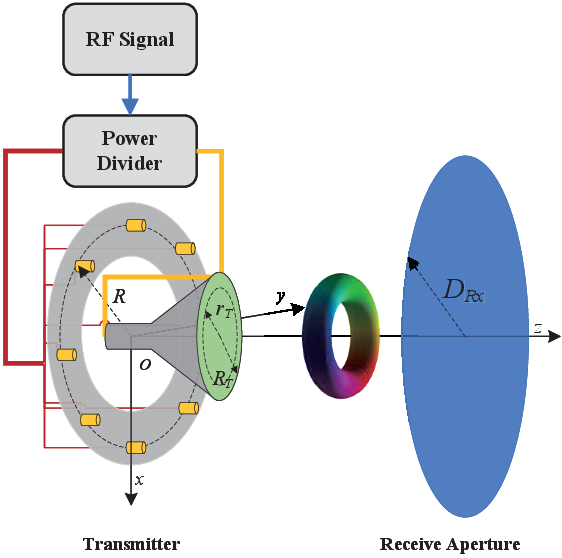}
\caption{System architecture of the toroidal-OAM hybrid beam launcher.}
\label{fig:system_schematic}
\end{figure}
We consider a point-to-point line-of-sight link as illustrated in Fig.~\ref{fig:system_schematic}. The transmitter integrates two coaxially aligned subsystems: (i) a coaxial horn antenna capable of generating toroidal electromagnetic pulses~\cite{wang2024observation}, whose far-field radiation exhibits inherent angular confinement and a strong longitudinal electric field component, and (ii) a UCA of $N$ elements at radius $R$ generating an OAM beam of integer topological charge $\ell \geq 1$ via progressive azimuthal phase excitation.  In local spherical coordinates, UCA element $n$ lies at azimuth $\varphi_n = 2\pi n/N$ ($n = 0, 1, \ldots, N-1$). Mode $\ell$ uses $w_n = e^{j\ell\varphi_n} = e^{j2\pi\ell n/N}$: equal amplitudes and phase increments $2\pi\ell/N$ radians complete $\ell$ full $2\pi$ cycles. The continuous-aperture cylindrical azimuthal front $e^{j\ell\varphi}$ is an eigenfunction of $L_z = -j\partial/\partial\varphi$ with eigenvalue $\ell$; geometry sets the polar-angle dependence.

The three transmitter stages are: \textbf{Stage~1}---a coaxial horn antenna, of the type demonstrated to generate toroidal electromagnetic pulses under broadband excitation~\cite{wang2024observation}, operates under narrowband continuous-wave (CW) excitation and is characterized by major radius $R_T$ and minor radius $r_T$. Under CW operation the full spatiotemporal pulse topology is not produced, but the horn's far-field radiation retains two characteristic features inherited from its toroidal-pulse-generating capability (Section~\ref{sec:toroidal_filter}): angular confinement modeled by the super-Gaussian envelope and a longitudinal electric field component $E_z$ from the radial polarization of the transverse electromagnetic (TEM) mode. \textbf{Stage~2}---the UCA independently generates the OAM beam with elements fed at equal amplitude and progressive phase $\phi_n = 2\pi\ell n/N$ (i.e., element $n$ is excited with $e^{j\ell\varphi_n}$, realizing the azimuthal phase front $e^{j\ell\varphi}$ of OAM mode $\ell$ in the continuous aperture limit); \textbf{Stage~3}---the two subsystems are integrated into a common radiating aperture via coaxial physical integration, with the horn occupying the central region and the UCA elements arranged as an annular ring at radius $R$ around it. At the aperture plane, a fraction of the UCA near-field radiation spatially overlaps with the horn's toroidal near-field region; this fraction, denoted by the geometric filling factor $\alpha \in [0,1]$, undergoes angular confinement by the toroidal envelope $T(\theta)$. The remaining fraction $(1-\alpha)$ propagates as conventional OAM radiation without toroidal modification. The horn independently contributes a longitudinal electric field component $E_z(\theta)$ via its radial polarization~\cite{wang2024observation}; the UCA is the sole source of OAM structure. An unequal Wilkinson power divider in the feed network splits the RF carrier power between the horn and UCA paths with split ratio $r_p = P_{\mathrm{horn}}/P_{\mathrm{UCA}}$, designed to satisfy $\alpha = \sqrt{r_p}/(1+\sqrt{r_p})$ (derived in Section~\ref{sec:analysis}) so that the geometric filling factor and the field amplitude weighting are governed by the same coefficient. Because both subsystems radiate from the same aperture plane, their electromagnetic fields undergo vector superposition in the near-field, forming a single hybrid wavefront that propagates to the far field as a unified toroidal-OAM beam. The longitudinal coupling $\beta = |E_z|/|E_{\perp}|$ is an inherent property of the horn's radial polarization and torus geometry. A finite-aperture receiver of diameter $D_{\mathrm{RX}}$ captures a portion of the angular power spectrum in the far field.


\section{Toroidal-OAM Hybrid Beam Analysis}
\label{sec:analysis}

This section develops the analytical framework of the toroidal-OAM hybrid beam. We begin with the conventional OAM-UCA far-field pattern, then introduce the toroidal radiation characteristics inherited from the coaxial horn, formulate the angular-domain hybrid beam model, derive the link budget and communication metrics, analyze the mode purity, and establish the physical parameter extraction relationships that connect the model to hardware-level quantities.

\subsection{Conventional OAM-UCA Far-Field Pattern}

For an $N$-element UCA in the $xy$-plane with element positions $\mathbf{r}_n = R(\cos\phi_n\,\hat{\mathbf{x}} + \sin\phi_n\,\hat{\mathbf{y}})$, $\phi_n = 2\pi n/N$, and progressive phase $e^{i\ell\phi_n}$, the array factor in the $\phi=0$ cut is $\mathrm{AF}(\theta) = \sum_{n=0}^{N-1} e^{i\ell\phi_n} e^{ikR\sin\theta\cos(\phi_n)}$, where $k = 2\pi/\lambda$ is the free-space wavenumber. In the continuous-aperture limit ($N \to \infty$) and using the Jacobi-Anger expansion, this reduces to~\cite{edfors2012LoSMIMO, mohammadi2010orbital}:
\begin{equation}
\label{eq:oam_field}
E_{\mathrm{OAM}}(\theta) = J_{\ell}(kR\sin\theta),
\end{equation}
with far-field angular power $P_{\mathrm{OAM}}(\theta) = |J_{\ell}(kR\sin\theta)|^2$. For $\ell > 0$, $J_{\ell}(0) = 0$, producing the characteristic donut-shaped intensity profile. The divergence half-angle is set by the first non-zero maximum $x'_{\ell 1}$ of $|J_{\ell}(x)|$:
\begin{equation}
\label{eq:oam_div}
\theta_{\mathrm{div}}^{\mathrm{(OAM)}} = \arcsin\!\left(\frac{x'_{\ell 1}}{kR}\right), \quad x'_{\ell 1} \approx \ell + 0.8086\,\ell^{1/3}.
\end{equation}
For $kR = 10\pi$ ($R=5\lambda$, $f=3$~GHz), $\theta_{\mathrm{div}}^{\mathrm{(OAM)}} \in [3.36^\circ, 17.89^\circ]$ for $\ell = 1$--$8$. 
Equation~\eqref{eq:oam_div} distinguishes the divergence angle $\theta_{\mathrm{div}}^{\mathrm{(OAM)}}$, which locates the mode-$\ell$ annular intensity maximum, from the beamwidth, which measures that lobe's angular width. Both are governed by the aperture-diffraction scale $\lambda/(2R)$, which also sets the half-power beamwidth of a conventional array of diameter $2R$. Since $x'_{\ell 1} \approx \ell$ for large $\ell$~\cite{abramowitz1964handbook}, the small-angle divergence grows approximately as $\ell/(kR)$: increasing mode order moves the main lobe outward while its angular width remains at the aperture-diffraction scale. At distance $z$, the beam footprint is an annulus of radius $z\tan\theta_{\mathrm{div}}^{\mathrm{(OAM)}}$.
 We assume $N \geq 2\ell_{\max}+1 = 17$ throughout, satisfying the Nyquist condition for clean mode generation~\cite{edfors2012LoSMIMO}.

\subsection{Toroidal Pulse Features and Phenomenological Model}
\label{sec:toroidal_filter}


The multipole origin of toroidal electromagnetic pulses can be understood through the electromagnetic toroidal dipole moment, defined for a time-harmonic current distribution $\mathbf{J}(\mathbf{r}; \omega)$ as:
\begin{equation}
\label{eq:toroidal_moment}
\mathbf{T}(\omega) = \frac{1}{10c}\int \bigl[(\mathbf{r}\cdot\mathbf{J})\,\mathbf{r} - 2r^2\mathbf{J}\bigr]\,\mathrm{d}^{3}r,
\end{equation}
where $c$ is the speed of light. A non-zero $\mathbf{T}(\omega)$ indicates that the source current distribution contains toroidal multipole character. The coaxial horn antenna~\cite{wang2024observation}---a coaxial waveguide whose inner and outer conductors support poloidal current paths---realizes the canonical current configuration that produces a non-zero $\mathbf{T}(\omega)$. Under broadband pulsed excitation, the coherent superposition of toroidal multipole components across a wide frequency range yields the full toroidal electromagnetic pulse with its spatiotemporal topology (nested field-line toroidal surfaces, skyrmionic textures, space-time nonseparability)~\cite{hellwarth1996focused, zdagkas2022observation, wang2024observation}.

In this work, the horn's far-field radiation under narrowband CW excitation retains two characteristic features attributed to the same poloidal current geometry that enables toroidal pulse generation. We model these features as:

\begin{enumerate}
  \item \textbf{Angular confinement}: The coaxial horn's far-field radiation exhibits a characteristically tight angular spread, a feature inherited from the toroidal pulse's spatial localization. This confinement is modeled by a super-Gaussian angular envelope of order $m=4$ (sharper than Gaussian, smoother than a hard aperture):
  \begin{equation}
  \label{eq:toroidal_filter}
  T(\theta) = \exp\!\left[-\left(\frac{\sin\theta}{\sin\theta_c}\right)^{4}\right], \quad \theta_c = \frac{\pi}{\kappa \cdot kR},
  \end{equation}
  where $\kappa$ is the toroidal confinement parameter.  Physically, a larger $\kappa$ yields tighter angular confinement (a smaller critical angle $\theta_c$), and the quantitative geometry dependence is given by Eq.~\eqref{eq:kappa_geometry} below. 

  \item \textbf{Longitudinal field}: Toroidal pulses possess a strong longitudinal electric field component~\cite{wang2024observation}, a distinctive feature not present in conventional transverse electromagnetic beams. Under axisymmetric, narrowband, paraxial conditions, the longitudinal far-field component is well approximated by a zeroth-order Bessel profile (axisymmetric, maximum on-axis):
  \begin{equation}
  \label{eq:elong}
  E_z(\theta) = J_{0}\!\left(\frac{\kappa \cdot kR \cdot \sin\theta}{\pi}\right).
  \end{equation}
  $E_z(0)=1$ captures the non-zero on-axis energy, and the first zero occurs at $\sin\theta \approx 2.4\theta_c$, well beyond the angular range of interest. The $J_0$ profile is the axisymmetric solution to the scalar Helmholtz equation, consistent with the longitudinal field component derived from the measured radial field via Gauss's law in the microwave realization~\cite{wang2024observation}.  $E_z(\theta)$ is a normalized scalar proxy for the horn's polarization-projected longitudinal component: Gauss-law retrieval~\cite{wang2024observation} establishes its presence at finite measurement distance; CW far-field persistence is assumed below. The scalar superposition~\eqref{eq:hybrid} represents the polarization-matched receive projection.
\end{enumerate}

Together, $T(\theta)$ and $E_z(\theta)$ capture the key far-field signatures of the toroidal pulse: angular confinement and longitudinal field. Their parameters are set by the coaxial horn geometry and the experimental measurements of Wang~\emph{et al.}~\cite{wang2024observation}. Under narrowband CW excitation, these signatures persist as monochromatic manifestations of the same poloidal current geometry that produces toroidal pulses under broadband excitation.

 The toroidal-OAM hybrid beam analyzed in this paper is a monochromatic far-field beam: the complete spatiotemporal toroidal topology, including space-time nonseparability and the skyrmionic field textures, exists only under broadband pulsed excitation. Under narrowband CW operation, the horn contribution is modeled by the angular confinement and a longitudinal profile inherited from the same poloidal current geometry, and the persistence of the longitudinal profile under CW operation is the modeling assumption discussed below. The term ``toroidal-OAM hybrid beam'' refers to this origin.

 The two model components are motivated by the following physical arguments. The longitudinal component $E_z$ was retrieved from the measured radial field via Gauss's law in~\cite{wang2024observation}, and the propagation-invariance of the full toroidal pulse is reported in~\cite{wang2024propagation}. In the scalar paraxial approximation, the far field of an axisymmetric aperture field is its zero-order Hankel transform,
\begin{equation*}
E_z(\theta) = C \int_{0}^{\infty} E_{\rho}(\rho)\, J_0(k\rho\sin\theta)\,\rho\,\mathrm{d}\rho,
\end{equation*}
where $E_{\rho}(\rho)$ is the radial aperture field and $C$ a normalization constant. The poloidal current paths of the coaxial horn concentrate the radial aperture field into a ring-like distribution at a characteristic radius $\rho_{\mathrm{eff}}$, for which the Hankel transform evaluates to $J_0(k\rho_{\mathrm{eff}}\sin\theta)$; identifying $\rho_{\mathrm{eff}} = \kappa R/\pi$ recovers Eq.~\eqref{eq:elong}. This construction motivates the phenomenological profile of Eq.~\eqref{eq:elong}, with the persistence of the same longitudinal profile under CW operation as a modeling assumption of this paper. For the angular confinement, the far-field envelope of an apodized aperture is set by the aperture apodization through the Fourier-Bessel relation, and the super-Gaussian envelope of order $m=4$ in Eq.~\eqref{eq:toroidal_filter} is adopted as a phenomenological envelope sharper than a Gaussian ($m=2$) while remaining smooth, in qualitative agreement with the tight angular roll-off reported for the coaxial horn in~\cite{wang2024observation}. The exponent $m=4$ and the critical-angle mapping $\theta_c = \pi/(\kappa kR)$ are empirical modeling parameters, and the geometry dependence of $\kappa$ is given by Eq.~\eqref{eq:kappa_geometry} below.

\subsection{Toroidal-OAM Hybrid Beam Formulation}

\subsubsection{Angular-Domain Model}

The complete toroidal-OAM hybrid beam in the far field is constructed as a linear superposition of three contributions: the unmodified OAM-UCA radiation, an angularly filtered OAM component, and an additive longitudinal pedestal. The horn and UCA share a common aperture plane, and their fields superpose in the far field. The hybrid field is expressed as:
\begin{align}
\label{eq:hybrid}
E_{\mathrm{hyb}}(\theta) &= \bigl[1 - \alpha + \alpha T(\theta)\bigr]E_{\mathrm{OAM}}(\theta) + \alpha\beta E_z(\theta)\nonumber\\
&= (1-\alpha)E_{\mathrm{OAM}}(\theta) + \alpha E_{\mathrm{OAM}}(\theta)T(\theta) + \alpha\beta E_z(\theta),
\end{align}
where $\alpha \in [0, 1]$ is the geometric filling factor, denoting the fraction of the UCA near-field radiation that spatially overlaps with the horn's toroidal near-field region, and $\beta = 0.3$ is the longitudinal coupling strength (the inherent longitudinal-to-transverse field ratio $|E_z|/|E_{\perp}|$ of the coaxial horn). The effective angular filter $1 - \alpha + \alpha T(\theta)$ interpolates between unity (clear aperture, no toroidal modification) and the super-Gaussian confinement $T(\theta)$ (horn-overlap region). The additive term $\alpha\beta E_z(\theta)$ is the longitudinal pedestal contributed by the horn via its radial polarization; it is azimuthally symmetric (carrying no OAM) and weighted by the same filling factor $\alpha$ because the $E_z$ source is co-located with the horn aperture. The helical phase structure of the OAM beam originates solely from $E_{\mathrm{OAM}}(\theta)$, which means the UCA is the only OAM-carrying element in the system. The two limits are $\alpha = 0$ (conventional OAM, no toroidal modification) and $\alpha = 1$ (fully toroidal-OAM: the entire UCA radiation is angularly confined and the longitudinal pedestal is at full strength).  The factorized form in the first line explicitly separates the spatial-filtering action on the UCA field from the additive longitudinal pedestal, and the second line shows the equivalent expanded three-term superposition. The normalized power pattern is $P(\theta) = |E_{\mathrm{hyb}}(\theta)|^2 / \max_{\theta} |E_{\mathrm{hyb}}|^2$, with key metrics $\theta_{\mathrm{div}} = \arg\max_{\theta > 0}^{\text{first}} P(\theta)$ and $P(0) = P(\theta)|_{\theta=0}$ ; for center-peaked patterns whose global maximum lies on the axis, the divergence metric of Section~V-C tracks this on-axis peak position instead.

\subsubsection{Two-Dimensional Transverse Extension}

For the transverse $(x,y)$ plane at observation distance $z$, $\rho = \sqrt{x^2 + y^2}$, $\phi = \arctan(y/x)$, $\sin\theta = \rho/\sqrt{\rho^2 + z^2}$. The 2D complex field is constructed by applying the OAM azimuthal phase to the OAM-carrying part of the hybrid field:
\begin{equation}
\label{eq:field_2d}
E_{\mathrm{hyb}}(x,y) = \bigl[(1-\alpha+\alpha T(\theta))E_{\mathrm{OAM}}(\theta)\bigr]e^{i\ell\phi} + \alpha\beta E_z(\theta),
\end{equation}
 where $E_{\mathrm{OAM}}(\theta) = J_{\ell}(kR\sin\theta)$. Equation~\eqref{eq:field_2d} applies azimuthal phase only to the OAM component; the azimuthally symmetric longitudinal pedestal is OAM-free, consistent with radial horn polarization. The scalar model's componentwise radial and azimuthal separability suffices for subsequent power-capture and mode-purity metrics.

\begin{figure}[htbp]
\centering
\subfigure[Conventional OAM $\ell=3$.]{
\begin{minipage}{0.45\linewidth}
\centering
\includegraphics[scale=0.38]{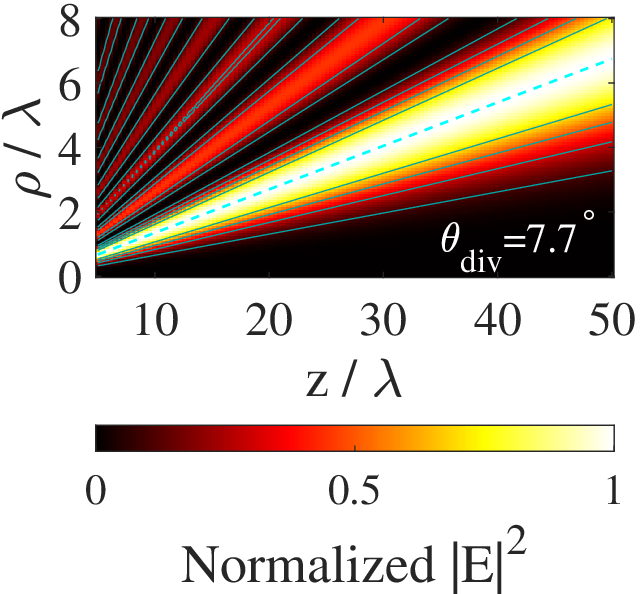}
\end{minipage}
}
\subfigure[Toroidal-OAM $\ell=3$, $\kappa=1.0$.]{
\begin{minipage}{0.45\linewidth}
\centering
\includegraphics[scale=0.38]{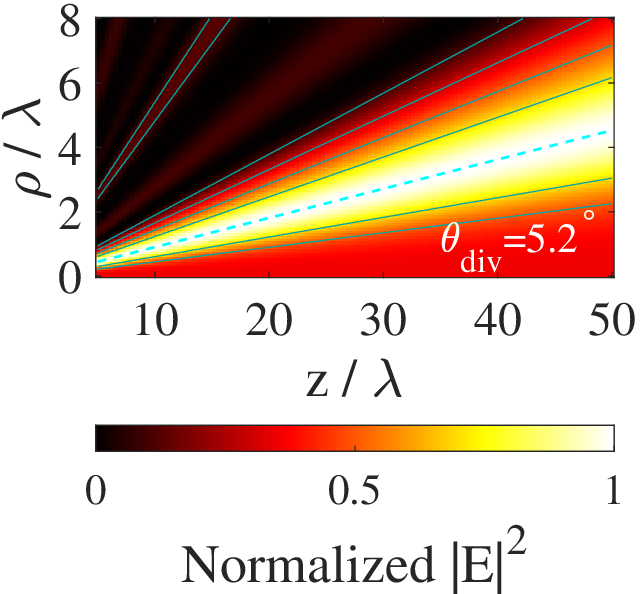}
\end{minipage}
}
\caption{Beam propagation in the $(\rho, z)$ plane for $\ell=3$ at configuration ($\kappa=1.0$, $\alpha=0.7$). (a) Conventional OAM: broad angular divergence with a dark central null. (b) Toroidal-OAM hybrid: reduced divergence (dashed cyan line), filled central axis from $E_z$, and tighter 0.1--0.7 intensity contours.}
\label{fig:propagation_plane}
\end{figure}

Fig.~\ref{fig:propagation_plane} visually demonstrates three key advantages: (i) the narrower angular spread between divergence reference lines (dashed cyan) confirms reduced far-field divergence; (ii) the bright axial trace at $\rho = 0$ in the toroidal-OAM case directly shows the $E_z$ null-filling effect; and (iii) the tighter intensity contours indicate improved spatial power concentration for finite-aperture reception. These advantages are quantified in Section~\ref{sec:results}.

\subsection{Link Budget and Communication Metrics}

The beam-level improvements translate to system-level communication performance through the power capture efficiency at a finite-aperture receiver. For a receiver of diameter $D_{\mathrm{RX}}$ at distance $z$, subtending half-angle $\theta_{\mathrm{RX}} = \arctan(D_{\mathrm{RX}}/2z)$. The power capture efficiency, denoted by $\eta_{\ell}(D_{\mathrm{RX}}, z)$, is given as follows:
\begin{equation}
\label{eq:power_capture}
\eta_{\ell}(D_{\mathrm{RX}}, z) = \frac{\int_{0}^{\theta_{\mathrm{RX}}} P_{\ell}(\theta)\,\sin\theta\,\mathrm{d}\theta}
                                          {\int_{0}^{\pi/2} P_{\ell}(\theta)\,\sin\theta\,\mathrm{d}\theta},
\end{equation}
where $P_{\ell}(\theta) = |E_{\mathrm{hyb}}(\theta)|^2$ for the hybrid and $|J_{\ell}(kR\sin\theta)|^2$ for conventional OAM. The received SNR for mode $\ell$ is:
\begin{equation}
\label{eq:snr}
\mathrm{SNR}_{\ell} = \frac{P_{\mathrm{TX}}\,G_{\mathrm{TX}}\,G_{\mathrm{RX}}\,\eta_{\ell}}{k_B T_0 B F \cdot L_{\mathrm{FS}}(z)},
\end{equation}
where $P_{\mathrm{TX}}$ is the transmit power, $G_{\mathrm{TX}}$ and $G_{\mathrm{RX}}$ are the transmit and receive antenna gains, $k_B$ is Boltzmann's constant, $T_0$ the noise temperature, $B$ the bandwidth, and $F$ the receiver noise figure; the free-space path loss is $L_{\mathrm{FS}}(z) = (4\pi z/\lambda)^2$. The SNR gain relative to conventional OAM, $\Delta\mathrm{SNR}_{\ell} = 10\log_{10}(\eta_{\ell}^{\mathrm{(Hyb)}}/\eta_{\ell}^{\mathrm{(OAM)}})$ [dB], is independent of all common-mode parameters. The corresponding range extension is:
\begin{equation}
\label{eq:range_ext}
\frac{z_{\mathrm{Hyb}}}{z_{\mathrm{OAM}}} = \sqrt{\frac{\eta_{\ell}^{\mathrm{(Hyb)}}}{\eta_{\ell}^{\mathrm{(OAM)}}}}.
\end{equation}

The power pattern $P_{\ell}(\theta) = |E_{\mathrm{hyb}}(\theta)|^2$ in~\eqref{eq:power_capture} evaluates the coherent superposition of~\eqref{eq:hybrid} at the constructive-interference azimuth of the OAM ring and longitudinal pedestal. Azimuthal averaging removes the interference term: $P_{\mathrm{avg}}(\theta) = \bigl|(1-\alpha+\alpha T)E_{\mathrm{OAM}}\bigr|^2 + |\alpha\beta E_z|^2$. Thus, Eq.~\eqref{eq:power_capture} is a normalized angular-cut concentration surrogate, with distance-dependent aperture contraction carried entirely by $\eta_{\ell}$. Absolute link budgets in Eq.~\eqref{eq:snr} and Section~\ref{sec:results} are illustrative; a fully energy-consistent absolute model remains future work. Fixed-distance SNR-gain and fairness comparisons use capture-efficiency ratios, whereas Table~\ref{tab:capacity} uses the stated nominal SNRs. With the averaged pattern, the per-mode SNR gains in Table~\ref{tab:link_performance} become $4.7$, $4.9$, $7.7$~dB at $z=100\lambda$ and $6.6$, $15.3$, $31.5$~dB at $z=300\lambda$; the capacity gains in Table~\ref{tab:capacity} remain of the same order, preserving the qualitative conclusions under either convention.

 The link-budget equivalence~\eqref{eq:range_ext} gives both schemes' distance ratio for equal received SNR with capture efficiencies fixed at one reference distance. Actual range requires full distance-dependent $\eta_{\ell}(D_{\mathrm{RX}}, z)$ in~\eqref{eq:power_capture}.

For $M$-mode OAM mode-division multiplexed (OAM-MDM), the power imbalance across modes is quantified by Jain's fairness index (JFI)~\cite{jain1984quantitative}:
\begin{equation}
\label{eq:jfi}
\mathcal{J}(\boldsymbol{\eta}) = \frac{(\sum_{\ell=1}^{M} \eta_{\ell})^2}{M \sum_{\ell=1}^{M} \eta_{\ell}^2} \in [1/M, 1].
\end{equation}

 Accounting for the per-mode power fraction $S_{\ell}$ and for residual modal cross-talk through the spectrum $\{S_n\}$, the effective signal-to-interference-plus-noise ratio (SINR) for mode $\ell$ is:
\begin{align}
\label{eq:sinr}
\mathrm{SINR}_{\ell} &= \frac{S_{\ell} \cdot \mathrm{SNR}_{\ell}}{1 + \sum_{n \neq \ell} L_{n,\ell} \cdot \mathrm{SNR}_n}.
\end{align}
 $L_{n,\ell}$ is mode $n$'s power fraction leaking into output harmonic $\ell$. Toroidal azimuthal symmetry ensures $L_{n,\ell} = 0$ for all $n \neq \ell$ when aligned. Mode-matched OAM outputs reject the zero-order pedestal, which enters only through numerator $S_{\ell}$: the intended-mode power fraction at the ring radius, a link-level per-mode power-efficiency surrogate.

\subsection{Mode Purity Analysis}

 The toroidal modification reshapes the azimuthal power distribution between the intended mode and the zero-order component. We quantify this via azimuthal Fourier decomposition of the complex field at the ring radius $r_{\mathrm{peak}}$. The mode purity spectrum can be defined as follows:
\begin{align}
\label{eq:mode_spectrum}
S_n &= \frac{|c_n|^2}{\sum_{m=0}^{\infty} |c_m|^2},
\end{align}
where
\begin{align}
\label{eq:mode_coeff}
c_n &= \frac{1}{2\pi}\int_{0}^{2\pi} E_{\mathrm{hyb}}(r_{\mathrm{peak}}, \phi)\, e^{-in\phi}\,\mathrm{d}\phi.
\end{align}
The intended mode has $S_{\ell} = 1$ for pure OAM. 
Azimuthally symmetric $T(\theta)$ and $E_z(\theta)$ do not mix OAM modes: Fourier coefficients $c_n$ vanish except at the intended mode $\ell$ and zero order. Thus, $S_{\ell}$ is the intended-mode power fraction at the ring radius, and $S_0 = 1 - S_{\ell}$ belongs to the symmetric non-OAM component. Under strong coupling, $S_{\ell} = 0.885$, $0.781$, and $0.541$ for $\ell = 1, 2, 3$ (Section~\ref{sec:results}), while computed inter-OAM-mode leakage is below $10^{-9}$. The fixed-DFT receiver's selected non-zero outputs reject the pedestal common to all modes, extracting each stream's OAM component without zero-order-induced crosstalk. Pedestal power is excluded from useful per-mode SINR; $S_{\ell}$ serves as the link-level surrogate for this efficiency loss. The spectrum $S_n$ enters Eq.~\eqref{eq:sinr} and the crosstalk-inclusive capacity of Eq.~\eqref{eq:capacity_xcorr}.

\subsection{Physical Parameter Extraction and Experimental Realizability}

The three dimensionless model parameters are linked to measurable transmitter hardware properties through the design relationships developed below, connecting the hybrid beam model to hardware-level quantities.


\subsubsection{Confinement parameter $\kappa$ from torus geometry}

The toroidal confinement parameter is linked to the torus major radius $R_T$ and minor radius $r_T$. In the far field, the diffraction-limited angular spread from the minor-radius aperture is $\Delta\theta_T \sim \lambda/r_T$, while the torus major radius sets the characteristic propagation direction. Equating the super-Gaussian critical angle $\theta_c$ with the diffraction limit yields:
\begin{equation}
\label{eq:kappa_geometry}
\kappa = \frac{\pi \cdot r_T}{\lambda} \cdot \frac{1}{kR} = \frac{r_T}{2R},
\end{equation}
where $R$ is the UCA radius. For the baseline $\kappa = 1.0$ and $R = 5\lambda$, this implies $r_T = 10\lambda$ minor radius. To achieve a desired $\kappa$, the coaxial horn should be designed with minor aperture radius $r_T = 2\kappa R$.  Since $r_T = 10\lambda$ exceeds the UCA radius, $\kappa = 1.0$ models an equivalent apodized aperture (effective minor radius $r_T = 10\lambda$); Section~\ref{sec:hfss}'s smaller-$\kappa$ geometry nests the horn tube ($r_T = 80$~mm, $R = 1.35\lambda$, $\kappa \approx 0.3$).


\subsubsection{Coupling coefficient $\alpha$ from aperture geometry and feed network design}

The coupling coefficient $\alpha$ has a dual physical interpretation. Geometrically, it is the aperture filling fraction $\alpha = A_{\mathrm{horn}}/A_{\mathrm{total}}$, where $A_{\mathrm{horn}}$ is the area of the central horn aperture and $A_{\mathrm{total}}$ is the total radiating aperture area. This filling fraction determines what portion of the UCA near-field radiation undergoes toroidal angular confinement $T(\theta)$. Electrically, the same coefficient can be related to the relative field amplitude between the two feed paths. At the feed network input, an unequal Wilkinson power divider splits the RF carrier with split ratio $r_p = P_{\mathrm{horn}}/P_{\mathrm{UCA}}$. Under the assumption of perfect impedance matching and equal path characteristics for both branches, the geometric filling fraction and the electrical field weighting can be aligned through the design relationship:
\begin{equation}
\label{eq:alpha_power}
\alpha = \frac{\sqrt{P_{\mathrm{horn}}}}{\sqrt{P_{\mathrm{horn}}} + \sqrt{P_{\mathrm{UCA}}}} = \frac{\sqrt{r_p}}{1 + \sqrt{r_p}}.
\end{equation}
The operational range $\alpha \in [0.3, 0.7]$ corresponds to $r_p \in [0.18, 5.4]$ and horn-to-total aperture area ratios of $30\%$--$70\%$, readily realizable using standard microstrip or waveguide power dividers with $\pm 0.5$~dB amplitude balance.  Equation~\eqref{eq:alpha_power} prescribes feed-network power splitting realizing field weight $\alpha$ throughout $\alpha \in [0.3, 0.7]$---a modeling design assumption, not a calibrated feed measurement. As $\alpha$ approaches $1$, UCA power tends to zero while Eq.~\eqref{eq:hybrid} retains the finite mathematical endpoint $(1-\alpha+\alpha T)J_{\ell} \to T J_{\ell}$. Their limiting divergence is a field-parameterization limitation, not a realizable feed state. In practice, $\alpha$ is set by the physical horn aperture diameter relative to the UCA ring radius at fabrication time, while the feed network power division is adjusted to match. Fine-tuning through variable attenuators in the feed path enables real-time optimization of the morphology--suppression tradeoff for different link conditions.

\subsubsection{Longitudinal coupling strength $\beta$ from horn field ratio}

The longitudinal coupling strength $\beta$ is the inherent longitudinal-to-transverse electric field ratio $|E_z|/|E_{\perp}|$ of the coaxial horn. The experimental characterization of the coaxial horn emitter in~\cite{wang2024observation}---where the longitudinal component $E_z$ was retrieved from the measured radial field via Gauss's law---confirms a non-negligible longitudinal field; the companion propagation analysis~\cite{wang2024propagation} further indicates that this longitudinal-to-transverse field ratio remains stable throughout the Fresnel and far-field regions. We adopt the conservative value $\beta = 0.3$ as the baseline, consistent with the qualitative experimental evidence. The horn's toroidal geometry (major radius $R_T$, minor radius $r_T$) sets both $\beta$ and $\kappa$ simultaneously: the radial polarization of the TEM mode in the coaxial horn directly produces the longitudinal component, with $|E_z|/|E_{\perp}| \propto r_T/R_T$ at the aperture plane.

\subsubsection{Model validation against limiting cases}

The limiting behaviors of the analytical model serve as consistency checks:
\begin{enumerate}
  \item \textbf{Pure OAM limit} ($\alpha = 0$): $E_{\mathrm{hyb}}(\theta) = J_{\ell}(kR\sin\theta)$, recovering the standard OAM-UCA far-field pattern verified experimentally~\cite{yan2014high, tamburini2012encoding}.
  \item \textbf{Hard aperture limit} ($\kappa \to \infty$, $m \to \infty$): $T(\theta) \to \mathrm{rect}(\theta/\theta_c)$, recovering the classical circular aperture diffraction pattern. The super-Gaussian $m = 4$ interpolates between Gaussian ($m = 2$, soft apodization) and hard-aperture limits.
  \item \textbf{Flying doughnut limit} ($\alpha = 1$, $\ell = 0$): $E_{\mathrm{hyb}}(\theta) = J_0(kR\sin\theta) \cdot T(\theta) + \beta J_0(\kappa kR\sin\theta/\pi)$, a Bessel-like beam with super-Gaussian apodization. We emphasize that the model~\eqref{eq:hybrid} describes a monochromatic far-field beam, whereas the experimental toroidal pulse is a broadband spatiotemporally nonseparable wavepacket; the connection is through the shared poloidal current geometry of the horn: under CW operation it produces angular confinement and longitudinal field at each frequency, while under broadband excitation the coherent superposition of these components yields the full pulse topology.
  \item \textbf{Reciprocity}: The far-field angular power pattern is reciprocal, i.e., $P_{\mathrm{TX}}(\theta) = P_{\mathrm{RX}}(\theta)$ for identical aperture distributions, ensuring divergence suppression benefits apply symmetrically to uplink and downlink.
\end{enumerate}


The hybrid beam~\eqref{eq:hybrid} captures three angular regimes (visualized in the far-field and 2D maps of Section~\ref{sec:results}): (i) \textbf{inner region} ($\theta < \theta_c$), where $T(\theta) \approx 1$ and the $E_z$ component fills the central null; (ii) \textbf{transition region} ($\theta \approx \theta_c$), where the super-Gaussian filter attenuates side-lobe radiation, manifesting as sharper ring edges; and (iii) \textbf{outer region} ($\theta \gg \theta_c$), where only the residual longitudinal field contributes. The longitudinal $J_0$ component, which is purely real and azimuthally symmetric, adds an in-phase pedestal that regularizes the central phase singularity. The design tradeoff is governed by $\kappa$: larger $\kappa$ reduces $\theta_c$, benefiting on-axis filling but risking main-lobe clipping for higher-order modes when $\theta_c$ falls below $\theta_{\mathrm{div}}^{\mathrm{(OAM)}}$. This defines the low-mode advantage regime, corresponding to $\ell \leq 3$ at $\kappa = 1.0$, $R = 5\lambda$, motivating mode-adaptive confinement for multi-mode operation.


\section{OAM-MDM Channel Capacity Analysis}
\label{sec:capacity}

The beam-level improvements of the toroidal-OAM hybrid---divergence suppression, central null filling, and the resulting power capture enhancement---translate directly to information-theoretic capacity gains in OAM mode-division multiplexing (MDM) systems. This section formalizes this connection by constructing the OAM channel matrix, deriving the ergodic capacity under both ideal and cross-talk-impaired conditions, and quantifying the capacity enhancement attributable to the hybrid beam. The derivation links the channel capacity to the power capture efficiency $\eta_{\ell}$ and mode purity spectrum $\{S_n\}$ developed in Section~\ref{sec:analysis}.

\subsection{OAM Channel Matrix}

Consider a transmit UCA of $N_t$ elements at radius $R_t$ and a receive UCA of $N_r$ elements at radius $R_r$, separated by distance $z$. The free-space channel matrix between individual array elements, denoted by $\mathbf{G} \in \mathbb{C}^{N_r \times N_t}$, has entries given by the Friis transmission formula:
\begin{equation}
\label{eq:G_matrix}
G_{n_r, n_t} = \frac{\lambda}{4\pi d_{n_r,n_t}} e^{-j k d_{n_r,n_t}},
\end{equation}
where $d_{n_r,n_t}$ is the distance between the $n_t$-th transmit element and the $n_r$-th receive element. OAM mode multiplexing and demultiplexing are realized via spatial discrete Fourier transform (DFT) beamforming matrices $\mathbf{W}_T \in \mathbb{C}^{N_t \times M}$ and $\mathbf{W}_R \in \mathbb{C}^{N_r \times M}$, with entries $[\mathbf{W}_T]_{n_t, \ell} = \frac{1}{\sqrt{N_t}} e^{j2\pi \ell (n_t-1)/N_t}$ and $[\mathbf{W}_R]_{n_r, \ell} = \frac{1}{\sqrt{N_r}} e^{j2\pi \ell (n_r-1)/N_r}$. The effective OAM channel matrix is:
\begin{equation}
\label{eq:H_OAM}
\mathbf{H}_{\mathrm{OAM}} = \mathbf{W}_R^H \mathbf{G} \mathbf{W}_T \in \mathbb{C}^{M \times M}.
\end{equation}

 For an ideal far-field free-space OAM-MDM link, (i)~coaxial, perfectly aligned UCAs with $N_t = N_r = N$ give $d_{n_r, n_t}$ in~\eqref{eq:G_matrix} depending only on $(n_t - n_r) \bmod N$, hence circulant $\mathbf{G}$. (ii)~DFT diagonalization makes $\mathbf{H}_{\mathrm{OAM}} = \mathbf{W}_R^H \mathbf{G} \mathbf{W}_T$ exactly diagonal, giving aligned-UCA OAM/LOS-MIMO equivalence under DFT beamforming~\cite{edfors2012LoSMIMO, mohammadi2010orbital}. (iii)~Per-mode gains are taken proportional to $\sqrt{\eta_{\ell}}$ in~\eqref{eq:H_diag}, phenomenologically mapping coherent UCA gains to the far-field angular-cut concentration surrogate in~\eqref{eq:power_capture}.  The aligned link's azimuthally symmetric toroidal modification redistributes power into the zero-order azimuthal component without off-diagonal OAM entries. The spectrum's~\eqref{eq:mode_spectrum} intended-mode power fraction $S_{\ell}$ enters SINR~\eqref{eq:sinr} and crosstalk-inclusive capacity~\eqref{eq:capacity_xcorr}. The per-mode channel gain is proportional to the square root of the power capture efficiency:
\begin{equation}
\label{eq:H_diag}
|H_{\ell,\ell}|^2 \propto \eta_{\ell}(D_{\mathrm{RX}}, z),
\end{equation}
where $\eta_{\ell}$ is given by~\eqref{eq:power_capture}.  Azimuthal symmetry nulls inter-mode leakage coefficients~\eqref{eq:sinr} in the aligned link. Off-diagonal per-mode spectral entries~\eqref{eq:mode_spectrum} quantify any leakage; intended-mode power fraction $S_{\ell}$ is the link-level per-mode power-efficiency factor.

\subsection{Ergodic Channel Capacity}

Under equal power allocation across $M$ OAM modes with total transmit power $P$, the ergodic channel capacity with perfect channel state information at the receiver is:
\begin{equation}
\label{eq:capacity}
C = \log_2 \det\!\left(\mathbf{I}_M + \frac{P}{M N_0} \mathbf{H}_{\mathrm{OAM}} \mathbf{H}_{\mathrm{OAM}}^H\right) \quad \text{[bps/Hz]},
\end{equation}
where $N_0 = k_B T_0 B F$ is the noise power. For the approximately diagonal OAM channel, the capacity decomposes into a sum of per-mode capacities:
\begin{equation}
\label{eq:capacity_diag}
C \approx \sum_{\ell=1}^{M} \log_2\!\left(1 + \frac{P}{M N_0} |H_{\ell,\ell}|^2\right) \quad \text{[bps/Hz]}.
\end{equation}

The channel state information (CSI) assumptions of~\eqref{eq:capacity} are perfect receiver knowledge of the effective OAM-domain channel (perfect CSIR), equal power across $M$ modes (no CSIT required), and a channel constant over the transmission block. Equation~\eqref{eq:capacity} gives the equal-power achievable rate; for deterministic LOS, its expectation equals the instantaneous value termed (ergodic) capacity. In the ideal aligned link, $\mathbf{H}_{\mathrm{OAM}}$ is exactly diagonal (as derived above), and geometry-determined fixed DFT matrices separate modes without estimating a general $N \times N$ MIMO channel. Per-mode pilots estimate only the $M$ gains $h_{\ell}$, whose magnitudes scale with $\sqrt{\eta_{\ell}}$ in~\eqref{eq:H_diag}. Under misalignment or scattering, off-diagonal crosstalk requires MIMO-style estimation, e.g., transmitting mode pilots, estimating $\mathbf{G}$, and forming $\hat{\mathbf{H}}_{\mathrm{OAM}} = \mathbf{W}_R^H \hat{\mathbf{G}} \mathbf{W}_T$ for linear/minimum mean squared error (MMSE) detection. Estimation errors leave residual inter-mode interference and reduce SINR; these effects are discussed qualitatively, not quantified, and Section~\ref{sec:results} assumes perfect CSIR. For ideal diagonal channels, both schemes use the same DFT demultiplexing structure~\cite{lyu2026eit}; identical per-mode pilot structures are a design assumption.

Under modal cross-talk from the toroidal modification, the effective SINR from~\eqref{eq:sinr} replaces the SNR in the per-mode capacity:
\begin{equation}
\label{eq:capacity_xcorr}
C_{\mathrm{hyb}} = \sum_{\ell=1}^{M} \log_2\!\left(1 + \mathrm{SINR}_{\ell}\right) \quad \text{[bps/Hz]}.
\end{equation}

\subsection{Capacity Enhancement from Toroidal-OAM Hybrid}

The capacity gain of the toroidal-OAM hybrid over conventional OAM arises from two mechanisms. First, the divergence suppression increases $\eta_{\ell}$ for each mode, boosting $|H_{\ell,\ell}|^2$ and therefore the per-mode SNR. Second, the central null filling contributes to total received power and mode equalization of total captured power at large distances. OAM demultiplexing rejects the zero-order pedestal, which enters Eq.~\eqref{eq:sinr} through intended-mode power fraction $S_{\ell}$ rather than as an additional coherent diversity path per mode. The net capacity enhancement is:
\begin{equation}
\label{eq:capacity_gain}
\begin{aligned}
\Delta C &= C_{\mathrm{hyb}} - C_{\mathrm{OAM}} \\
&\quad= \sum_{\ell=1}^{M} \log_2\!\left(\frac{1 + \mathrm{SINR}_{\ell}^{\mathrm{(Hyb)}}}{1 + \mathrm{SNR}_{\ell}^{\mathrm{(OAM)}}}\right) \quad \text{[bps/Hz]}.
\end{aligned}
\end{equation}

For ideal diagonal channels ($S_{\ell} = 1$ and no inter-mode interference), with every included mode at high SNR and common absolute normalization, $\Delta C \approx 0.332 \sum_{\ell=1}^{M} \Delta\mathrm{SNR}_{\ell}^{\mathrm{(dB)}}$~[bps/Hz]; crosstalk-inclusive gains are evaluated numerically through Eq.~\eqref{eq:capacity_gain}, with hybrid capacities $8.14\%$, $5.30\%$, and $3.89\%$ below their ideal values at SNR$=20$, $30$, and $40$~dB, respectively. Table~\ref{tab:capacity} instead normalizes each scheme to its own peak ($\gamma_X = \gamma_0\,\eta_X/\max \eta_X$) and reports $\Delta C = C_{\mathrm{Hyb}}^{\mathrm{(ideal)}} - C_{\mathrm{OAM}}$, so this asymptotic expression is not directly applied there. The detailed derivation and its remainder bound are provided in our response to the Editor due to space constraints.
  Table~\ref{tab:capacity} lists $C_{\mathrm{OAM}}$, $C_{\mathrm{Hyb}}^{\mathrm{(ideal)}}$, and $C_{\mathrm{Hyb}}^{\mathrm{(XT)}}$ at each SNR; Fig.~\ref{fig:capacity} compares both schemes versus SNR and mode count; Eq.~\eqref{eq:capacity_gain} summarizes their difference.  Ring-based fraction $S_{\ell}$ does not replace the full receive model's aperture-integrated mode overlap (misalignment analysis, Section~\ref{sec:misalignment}); unification with link-level capacity remains future work.

Numerical evaluation of the capacity enhancement is presented in Section~\ref{sec:results}.


\section{Numerical Results and Analysis}
\label{sec:results}

This section presents a comprehensive numerical evaluation of the toroidal-OAM hybrid beam using the analytical model developed in Section~\ref{sec:analysis}. We examine the transverse-plane field distributions, divergence suppression, central null filling, far-field angular power patterns, communication link performance, and channel capacity. All results are computed under the strong-coupling optimized configuration ($\alpha=0.7$, $\kappa=1.0$) unless otherwise noted.

\subsection{Simulation Setup}

Default parameters (strong-coupling optimized): $f = 3$~GHz ($\lambda \approx 0.1$~m), $R = 5\lambda$ ($kR = 10\pi$), $\kappa = 1.0$, $\alpha = 0.7$, $\beta = 0.3$. The coupling coefficient $\alpha=0.7$ is selected via systematic parameter sweep to maximize toroidal-OAM advantage; a quantitative comparison with balanced coupling ($\alpha=0.5$) is provided in Table~\ref{tab:alpha_comparison}. All core numerical results in this section assume the optimized parameter set: $f = 3$~GHz, $R = 5\lambda$, $\kappa = 1.0$, $\alpha = 0.7$, $\beta = 0.3$, $D_{\mathrm{RX}} = 20\lambda$ unless otherwise specified. Angular computation uses $N_\theta = 7200$ points over $\theta \in (0, 60^\circ]$; transverse-plane computation uses a $201\times 201$ grid over $5\lambda \times 5\lambda$ at $z = 10\lambda$. OAM modes $\ell = 1$--$8$ are evaluated under loose ($\kappa = 0.5$), baseline ($\kappa = 1.0$), and tight ($\kappa = 2.0$) confinement. The $E_z$ longitudinal component preserves the full signed Bessel $J_0(\cdot)$ profile without magnitude rectification, ensuring correct cross-term interference in the hybrid field superposition. The divergence angle is computed using the main lobe method (first non-zero local maximum of the far-field power pattern).


\subsection{Two-Dimensional Transverse-Plane Field Analysis}

\begin{figure}[htbp]
\centering
\subfigure[Conventional OAM $\ell=1$.]{\includegraphics[scale=0.2]{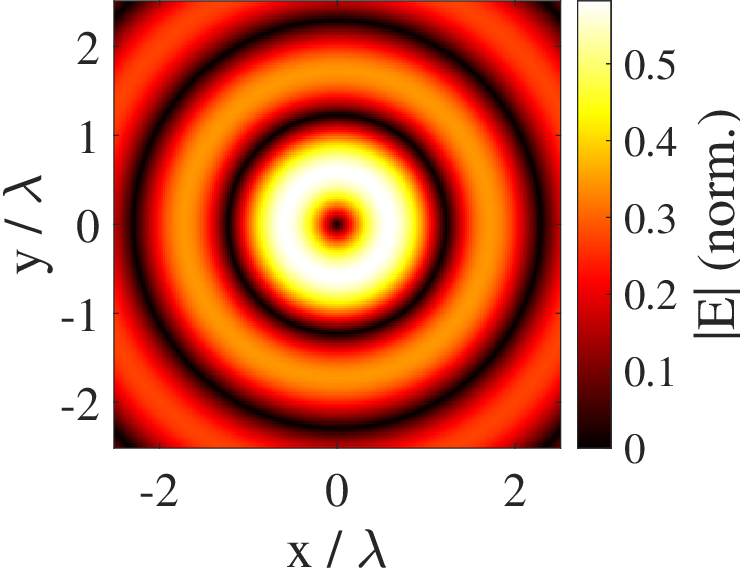}}
\subfigure[Conventional OAM $\ell=3$.]{\includegraphics[scale=0.2]{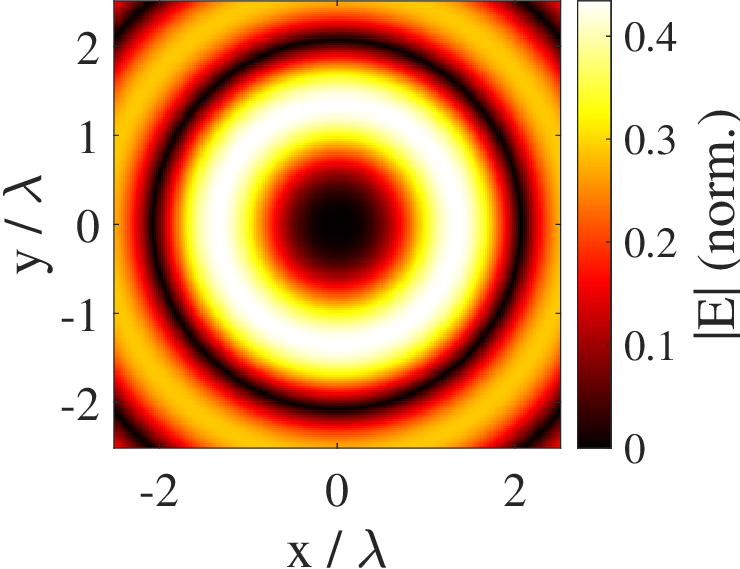}}
\subfigure[Conventional OAM $\ell=5$.]{\includegraphics[scale=0.2]{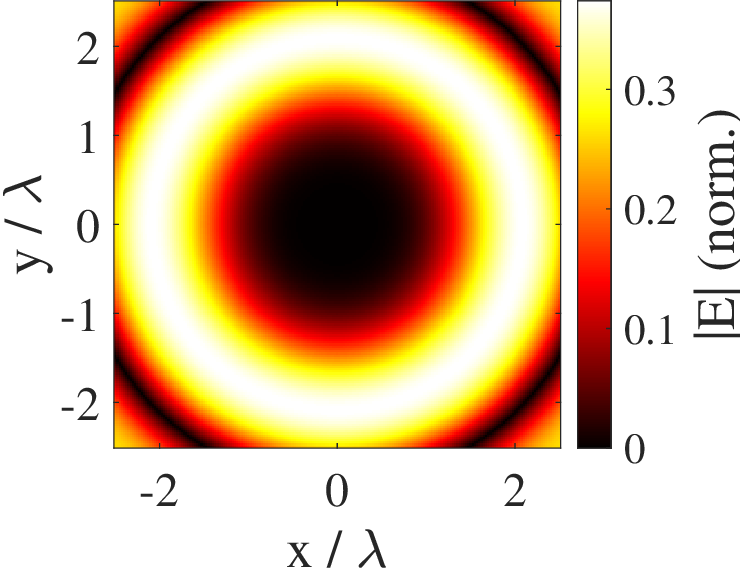}}\\
\subfigure[Toroidal-OAM $\ell=1$, $\kap=1.0$.]{\includegraphics[scale=0.2]{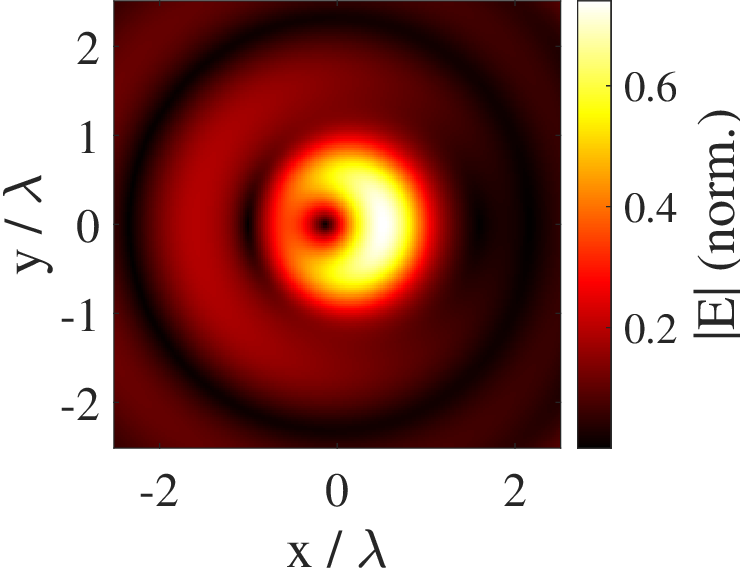}}
\subfigure[Toroidal-OAM $\ell=3$, $\kap=1.0$.]{\includegraphics[scale=0.2]{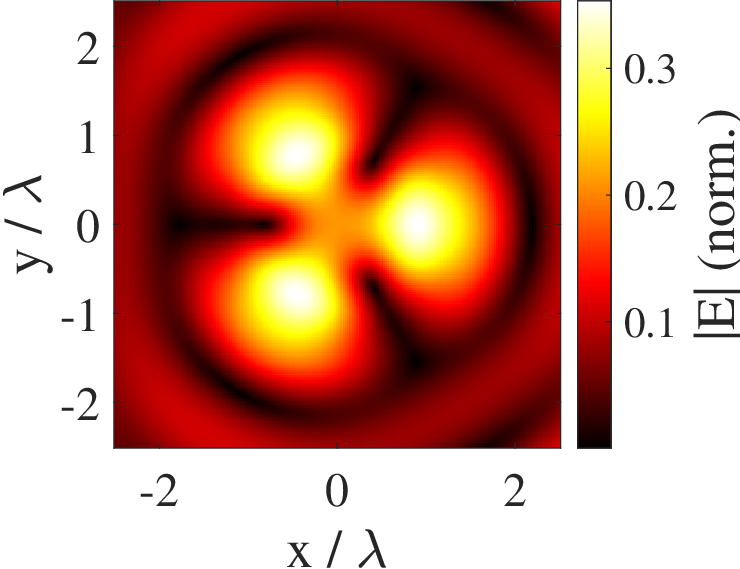}}
\subfigure[Toroidal-OAM $\ell=5$, $\kap=1.0$.]{\includegraphics[scale=0.2]{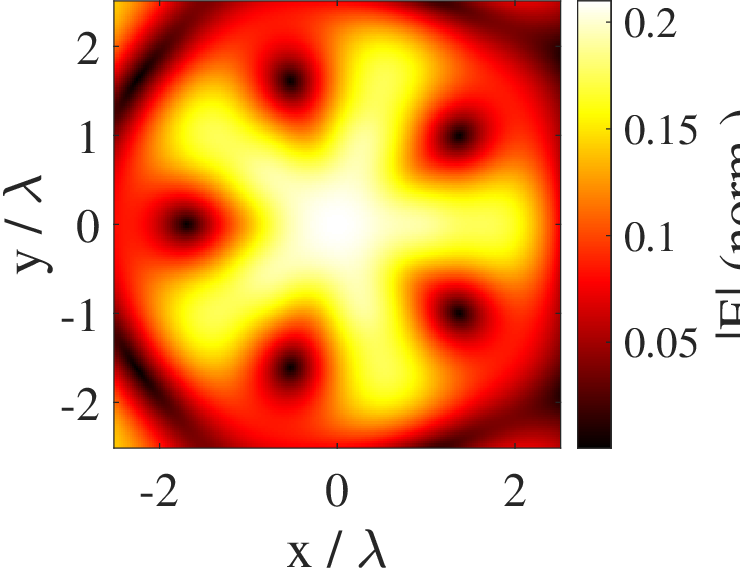}}
\caption{Normalized electric field amplitude $|E(x,y)|$ in the transverse plane at $z=10\lambda$ ($\kap=1.0$, $\alp=0.7$). Top row: conventional OAM-UCA beams for $\ell=1,3,5$. Bottom row: corresponding toroidal-OAM hybrid beams. Under strong coupling, the toroidal-OAM beams exhibit a visibly filled central region from the $E_z$ longitudinal component, an inward-shifted main ring (most pronounced at $\ell=3$), and an azimuthal modulation of the ring arising from the interference of the helical OAM component with the non-helical longitudinal pedestal, whose depth grows with the mode order. At $\ell=5$, the central $E_z$ pedestal becomes the dominant intensity feature, illustrating the transition from OAM-dominated to toroidal-dominated beam structure at higher mode indices.}
\label{fig:beam_amplitude}
\end{figure}

Fig.~\ref{fig:beam_amplitude} shows the transverse-plane amplitude $|E(x,y)|$ for conventional and toroidal-OAM beams at $\ell=1,3,5$ with $\alpha=0.7$. Three toroidal-OAM features are directly visible under strong coupling: (i) the main ring is shifted inward, by $8.7\%$ at $\ell=1$ and $33.3\%$ at $\ell=3$ (at $\ell=5$ the profile becomes center-dominated), consistent with the divergence reduction in Table~\ref{tab:divergence}; (ii) the central region is prominently filled by the longitudinal $E_z$ component, with the on-axis intensity at $z=10\lambda$ reaching $8.0\%$ of the global peak at $\ell=1$ and $35.3\%$ at $\ell=3$, and becoming the dominant feature at $\ell=5$ ($100\%$ of peak), matching the angular-model values in Table~\ref{tab:divergence}; and (iii) because the longitudinal pedestal carries no helical phase, its interference with the OAM ring produces an azimuthal modulation of the ring intensity whose depth grows with the mode order, the transverse-plane signature of the coherent superposition in Eq.~\eqref{eq:field_2d}. The progressive central brightening with $\ell$ reflects the growing dominance of the $\ell$-independent $E_z$ pedestal relative to the decreasing OAM peak intensity: at $\ell=5$, the main lobe angle substantially exceeds the toroidal critical angle $\theta_c \approx 5.7^\circ$, causing the super-Gaussian filter to strongly attenuate the OAM component while the $E_z$ term fundamentally transforms the beam profile from ring-shaped to center-peaked. This $\ell$-dependent structural evolution is a distinctive signature of the strong-coupling regime ($\alpha \geq 0.7$) and differs qualitatively from weaker coupling where the OAM ring remains dominant across all $\ell$.

\begin{figure}[htbp]
\centering
\subfigure[Conventional OAM $\ell=1$.]{\includegraphics[scale=0.2]{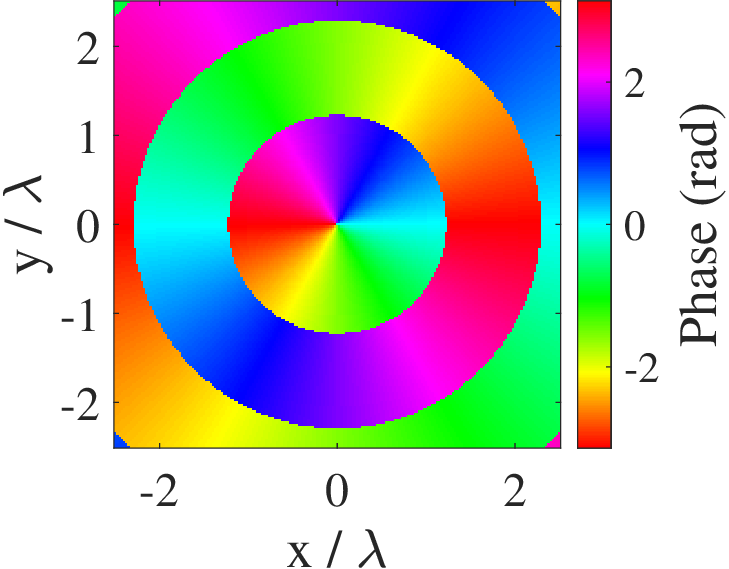}}
\subfigure[Conventional OAM $\ell=3$.]{\includegraphics[scale=0.2]{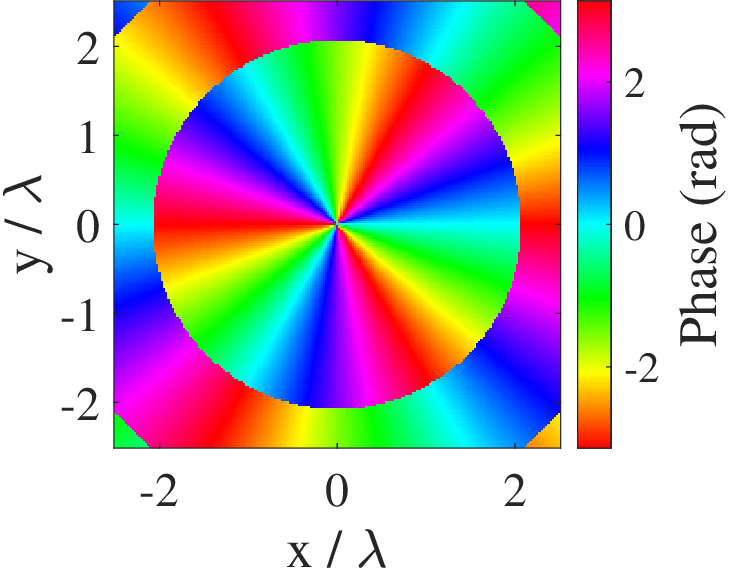}}
\subfigure[Conventional OAM $\ell=5$.]{\includegraphics[scale=0.2]{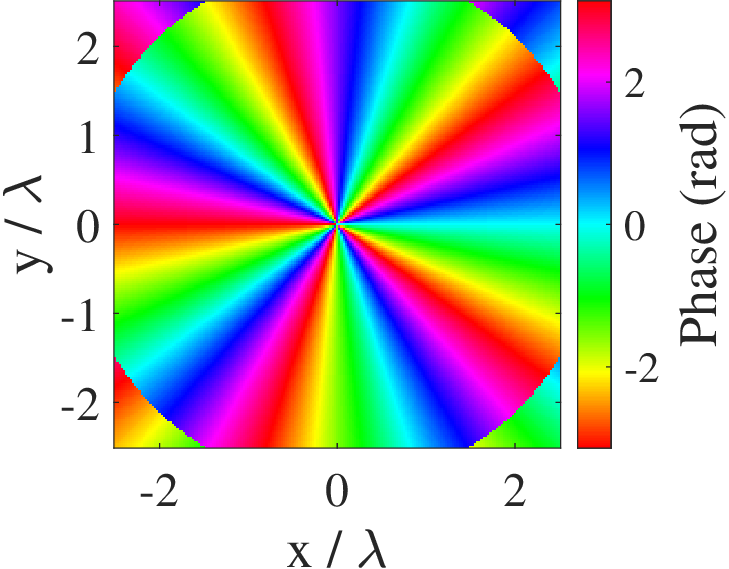}}\\
\subfigure[Toroidal-OAM $\ell=1$, $\kap=1.0$.]{\includegraphics[scale=0.2]{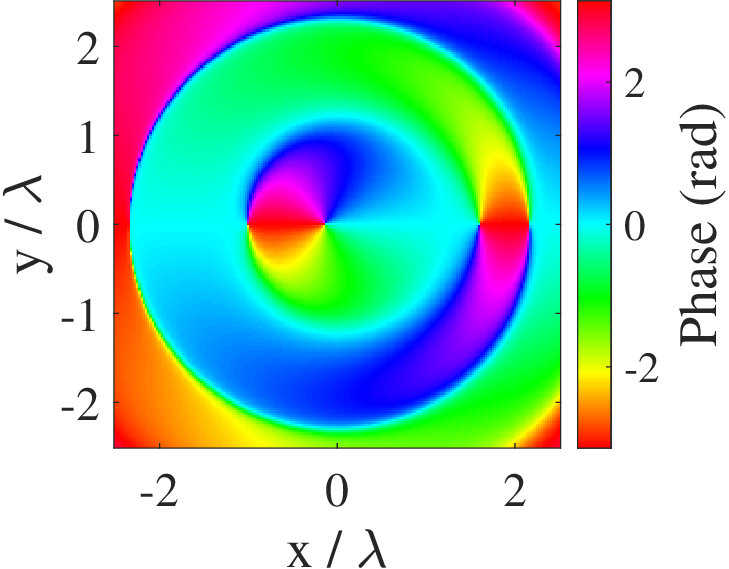}}
\subfigure[Toroidal-OAM $\ell=3$, $\kap=1.0$.]{\includegraphics[scale=0.2]{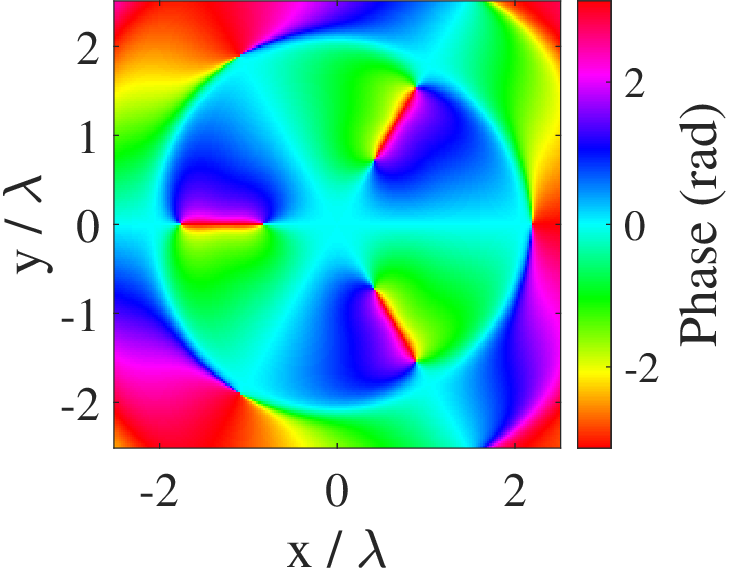}}
\subfigure[Toroidal-OAM $\ell=5$, $\kap=1.0$.]{\includegraphics[scale=0.2]{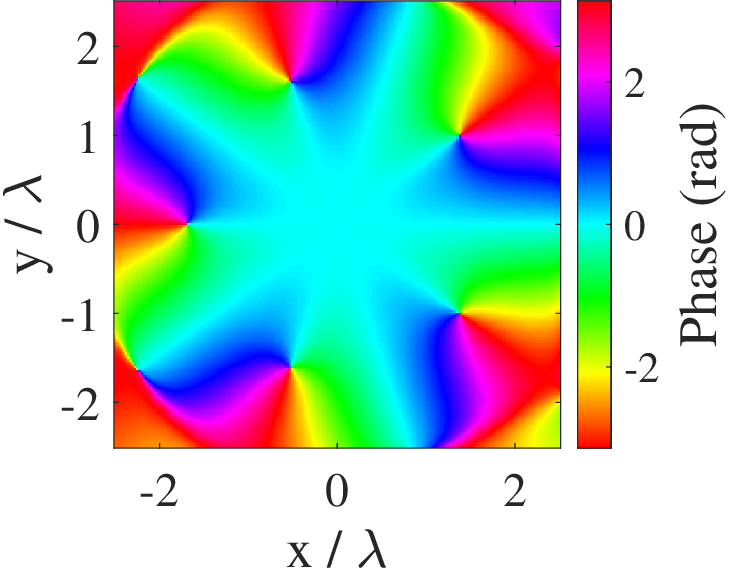}}
\caption{ Phase distribution $\arg[E(x,y)]$ in the transverse plane, comparing conventional OAM (top) and toroidal-OAM (bottom) for $\ell=1,3,5$. The conventional OAM phase exhibits the characteristic $\ell$-arm spiral with a central singularity (undefined phase at zero amplitude). The toroidal-OAM phase retains the $\ell$-fold azimuthal structure but shows regularization near the origin: the longitudinal $E_z$ component provides a real positive field that anchors the phase to zero at $\rho=0$, directly confirming the central null filling mechanism. Consistent with Eq.~\eqref{eq:field_2d}, the non-helical pedestal localizes the $\ell$ phase windings around the OAM ring, with minor amplitude-dependent phase deviations at the ring periphery where the super-Gaussian filter attenuates the OAM component.}
\label{fig:beam_phase}
\end{figure}

Fig.~\ref{fig:beam_phase} presents the corresponding phase distributions. The conventional OAM phase follows $\arg[E] \approx \ell\phi$ modulo $2\pi$, with $\ell$ azimuthal phase wrap lines converging at the central singularity. The toroidal-OAM phase retains this $\ell$-arm spiral structure, confirming that the toroidal modification does not alter the OAM mode character, but with a key difference: at the center, the phase becomes well-defined because the longitudinal $E_z$ component anchors the phase to zero at $\rho=0$---the phase-domain signature of the null filling mechanism. The electric field vectors remain dominantly azimuthal ($|E_\phi|/|E_\rho| > 20$), confirming that the OAM vortex topology is preserved.

\subsection{Divergence Suppression}

\begin{table}[htbp]
\centering
\caption{Divergence angle comparison: conventional OAM-UCA vs.\ toroidal-OAM hybrid beam ($\kap=1.0$, $\alp=0.7$, $R=5\lambda$, $f=3$~GHz). Ratio $<1$ indicates reduction. $R_{\mathrm{ring}} = P(\thetadiv)/P(0)$: values $>1$ confirm OAM ring dominance even at strong coupling.}
\label{tab:divergence}
\begin{tabular}{c c c c c c}
\toprule
$\ell$ & $\thetadiv^{\mathrm{(OAM)}}$ & $\thetadiv^{\mathrm{(Hyb)}}$ & Ratio & $\onaxis$ & $R_{\mathrm{ring}}$ \\
\midrule
1 & $3.36^\circ$ & $2.97^\circ$ & 0.883 & 0.080 & 12.44 \\
2 & $5.58^\circ$ & $4.28^\circ$ & 0.768 & 0.160 & 6.23 \\
3 & $7.68^\circ$ & $5.18^\circ$ & 0.674 & 0.353 & 2.83 \\
4 & $9.74^\circ$ & $5.71^\circ$ & 0.586 & 0.736 & 1.36 \\
5 & $11.78^\circ$ & $0.00^\circ$ & 0.000 & 1.000 & 1.00 \\
6 & $13.81^\circ$ & $0.00^\circ$ & 0.000 & 1.000 & 1.00 \\
7 & $15.84^\circ$ & $0.00^\circ$ & 0.000 & 1.000 & 1.00 \\
8 & $17.89^\circ$ & $0.00^\circ$ & 0.000 & 1.000 & 1.00 \\
\bottomrule
\end{tabular}
\end{table}

Table~\ref{tab:divergence} quantifies divergence suppression under the strong-coupling configuration ($\alpha=0.7$). For $\ell=1$--$4$, toroidal-OAM achieves a progressive divergence reduction from $11.7\%$ to $41.4\%$, with the benefit increasing with $\ell$; for $\ell \geq 5$, the hybrid beam becomes fully center-peaked ($\theta_{\mathrm{div}}^{\mathrm{(Hyb)}} = 0^\circ$, ratio $= 0$). The mean ratio across $\ell=1$--$8$ is 0.364. The morphology ratio $R_{\mathrm{ring}} = P(\thetadiv)/P(0)$ (rightmost column) provides a quantitative test of the beam structure: $R_{\mathrm{ring}} > 1$ indicates ring dominance, while $R_{\mathrm{ring}} = 1$ signals that the center is the global maximum. Under strong coupling, $R_{\mathrm{ring}}$ decreases from $12.44$ at $\ell=1$ to $1.00$ at $\ell \geq 5$, marking the morphological boundary from ring-dominated to center-dominated beam structure. This mode-dependent behavior follows from the alignment between the OAM main lobe and the toroidal critical angle $\theta_c = \pi/(\kappa kR) \approx 5.7^\circ$: lower-order modes ($\ell \leq 2$) lie within or near $\theta_c$, yielding moderate filtering; intermediate modes ($\ell = 3$--$4$) straddle $\theta_c$, producing the strongest relative suppression; and higher-order modes ($\ell \geq 5$) substantially exceed $\theta_c$, where the filter effectively suppresses the OAM ring, leaving the $E_z$ pedestal as the sole dominant feature. The on-axis field $\onaxis$ increases from $8.0\%$ ($\ell=1$) to $100\%$ ($\ell \geq 5$) of peak intensity, directly reflecting the growing dominance of the $\alpha$-scaled $E_z$ contribution.

 The far-field model in Eqs.~\eqref{eq:toroidal_filter}--\eqref{eq:hybrid} gives two-step divergence reduction. First, super-Gaussian $T(\theta)$ decreases with $\theta$, chiefly near $\theta_c$ of~\eqref{eq:toroidal_filter}, attenuating the annular lobe's outer wing and shifting the composite pattern's first maximum inward. Second, the additive longitudinal pedestal $\alpha\beta E_z$ supplies an $\ell$-independent on-axis component, pulling intensity further inward. The hybrid divergence half-angle, the modified power pattern's first positive maximum, is $2.97^\circ$, $4.28^\circ$, $5.18^\circ$, and $5.71^\circ$ for $\ell=1$--$4$ in Table~\ref{tab:divergence}, versus conventional OAM's $3.36^\circ$--$9.74^\circ$. These ring maxima lie at or below $\theta_c$; $5.71^\circ$ at $\ell=4$ essentially equals $\theta_c \approx 5.73^\circ$. Lobes originally far beyond $\theta_c$ are suppressed toward it; $\ell \geq 5$ modes become pedestal-dominated and center-peaked. For fixed mode, $\kappa$, $\alpha$, and $\beta$, the suppression ratio depends on dimensionless $kR$, as quantified below.

 Under uniform phasing, a same-aperture rectangular array has broadside half-power beamwidth scale $\lambda/(2R)$: $R = 5\lambda$ gives $\lambda/(2R) \approx 5.73^\circ$, comparable to $\theta_c$. Programmable apertures support distinct LOS-MIMO modes~\cite{edfors2012LoSMIMO}. The hybrid reduces per-mode OAM divergence toward this diffraction scale; the apertures serve different roles rather than competing on one link metric.

\begin{figure}[htbp]
\centering
\subfigure[Divergence angles under $\kappa = 0.5, 1.0, 2.0$.]{
\begin{minipage}{0.45\linewidth}
\centering
\includegraphics[scale=0.28]{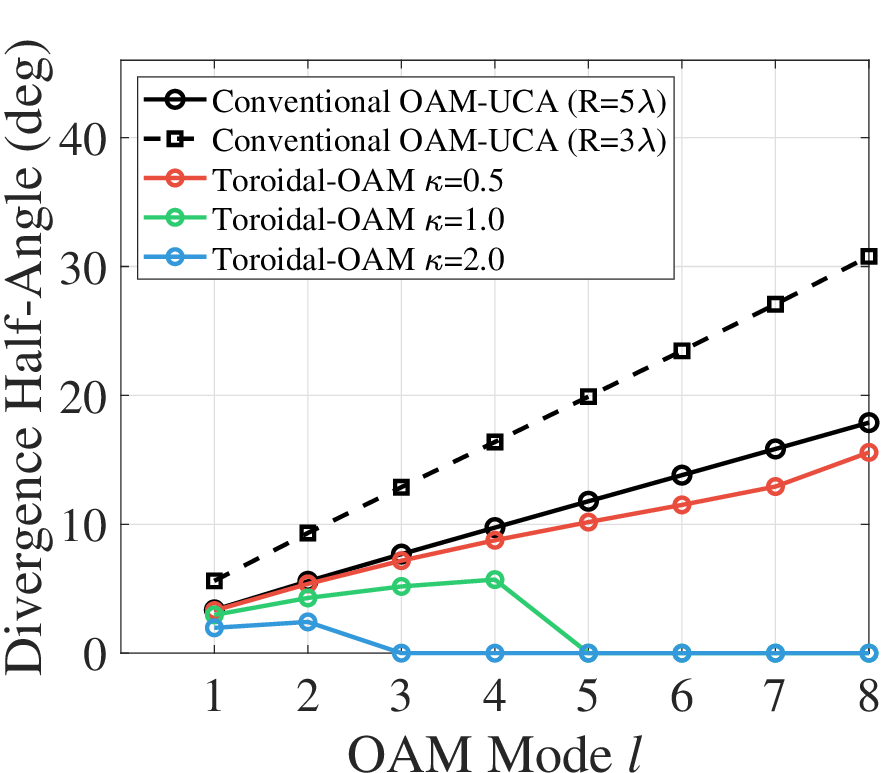}
\end{minipage}
}
\subfigure[Suppression ratio: $\kappa=2.0$ best for $\ell=1$--$2$, $\kappa=1.0$ for $\ell=3$--$4$, $\kappa=0.5$ for $\ell \geq 5$.]{
\begin{minipage}{0.45\linewidth}
\centering
\includegraphics[scale=0.28]{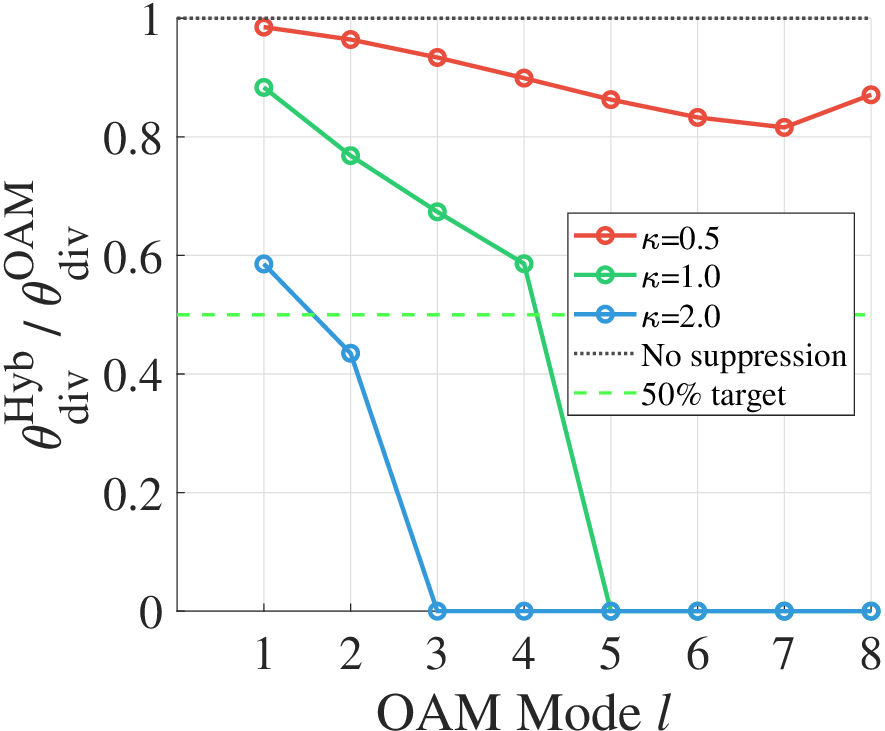}
\end{minipage}
}
\caption{Divergence analysis under strong coupling ($\alp=0.7$). (a) Divergence half-angle vs.\ $\ell$ at three $\kappa$ values: toroidal-OAM reduces divergence for all $\ell=1$--$8$, with $\kappa=1.0$ providing the best overall balance. (b) Suppression ratio: optimal $\kappa$ is mode-dependent.}
\label{fig:divergence_analysis}
\end{figure}

Fig.~\ref{fig:divergence_analysis} extends the divergence analysis under $\alpha=0.7$. The optimal $\kappa$ is mode-dependent: tighter confinement ($\kappa=2.0$) achieves the best single-mode results with $40.4\%$ reduction for $\ell=1$ ($3.36^\circ \to 2.00^\circ$) and $54.3\%$ for $\ell=2$ ($5.58^\circ \to 2.55^\circ$); baseline confinement ($\kappa=1.0$) is optimal for $\ell=3$--$4$; and looser confinement ($\kappa=0.5$) benefits $\ell \geq 5$. All modes $\ell=1$--$8$ show ratio $<1$ under $\kappa=1.0$, confirming universal divergence suppression, with the benefit extending meaningfully to $\ell=4$ which at balanced coupling shows only marginal improvement.

The suppression ratio is strictly independent of carrier frequency ($1$--$6$~GHz sweep) and normalized aperture ($R/\lambda = 2$--$10$ sweep), governed entirely by the dimensionless parameters $kR$, $\kappa$, and $\ell$. This frequency and aperture independence indicates that the toroidal-OAM mechanism follows a fundamental scaling law: the suppression performance depends only on normalized parameters, not on absolute physical dimensions, making the technique inherently broadband and scalable to different aperture sizes.  Within the scalar model, azimuthally symmetric $T(\theta)$ and $E_z$ preclude toroidal-induced inter-OAM-mode leakage at any wavelength (Section~\ref{sec:analysis}), while the intended-mode power fraction $S_{\ell}$ (Eq.~\eqref{eq:mode_spectrum}) depends only on $kR$ for fixed mode set, $\kappa$, $\alpha$, and $\beta$. Wavelength-scaled apertures thus preserve the modeled spectrum and $S_{\ell}$ at fixed $kR$, consistently with the frequency-invariance sweep.

For OAM-MDM, the divergence spread across modes is quantified by the coefficient of variation (CV): $\mathrm{CV} = 47.2\%$ (conventional) vs.\ $56.9\%$ (toroidal-OAM at $\kappa=1.0$, $\alpha=0.7$). The CV increases because strong coupling produces a sharp transition at $\ell=5$: modes $\ell=1$--$4$ retain OAM ring structure with reduced divergence, while modes $\ell \geq 5$ become fully center-peaked ($\theta_{\mathrm{div}} = 0^\circ$), creating a bimodal divergence distribution. Mode-adaptive $\kappa$ per mode group would be required for meaningful equalization, though the multi-mode fairness improvement (Table~\ref{tab:fairness}) partially compensates for this effect.  Equal-divergence designs use mode-dependent radii of concentric UCAs~\cite{lin2021equaldivergence}; the hybrid uses one shared aperture and angular filter. Mode-adaptive $\kappa$ for residual bimodality at $\ell \geq 5$ remains prospective; simultaneous per-mode $\kappa$ with one shared fixed filter is undemonstrated.

\subsection{Central Null Filling}

The conventional OAM beam has exactly $\onaxis = |J_{\ell}(0)|^2 = 0$ for all $\ell > 0$ due to the phase singularity at the beam axis. The toroidal-OAM hybrid fills this null through the longitudinal $E_z$ component, which follows the axisymmetric $J_0$ profile (Eq.~\ref{eq:elong}) with maximum on-axis. Under strong coupling ($\alpha=0.7$), the on-axis field increases from $8.0\%$ ($\ell=1$) to $100\%$ ($\ell \geq 5$) of the peak intensity (Table~\ref{tab:divergence}). The increase with $\ell$ follows from the decreasing OAM peak intensity as the main lobe broadens, while the $\ell$-independent $E_z$ pedestal magnitude remains constant. A parameter sweep across $\kappa$ at $\ell=4$ (not shown) confirms that the on-axis field grows monotonically with $\kappa$ (from $3.1\%$ at $\kappa=0.2$ to $54.0\%$ at $\kappa=2.0$) and the divergence ratio reaches a formal minimum of $0.323$ at $\kappa \approx 1.63$; however, this minimum corresponds to mode structure transformation (the detected ``peak'' is the central $E_z$ lobe, not the OAM ring) rather than genuine divergence suppression. The practical design range is $\kappa \in [0.5, 1.2]$ where the OAM mode character is preserved.

For practical receivers, a non-zero on-axis field simplifies alignment: an on-axis detector element can sense the presence of any OAM mode without mode-matched annular receivers. For $\ell \leq 3$, most communication power still resides in the annular main lobe, while for $\ell \geq 4$, the central lobe becomes the dominant power-bearing feature. The null filling serves as both an alignment and detection enabler at all modes and a supplementary power delivery channel at higher mode indices, though mode-matched reception remains preferred for maximizing link budget.

\subsection{Far-Field Angular Power Patterns}

\begin{figure}[htbp]
\centering
\includegraphics[scale=0.35]{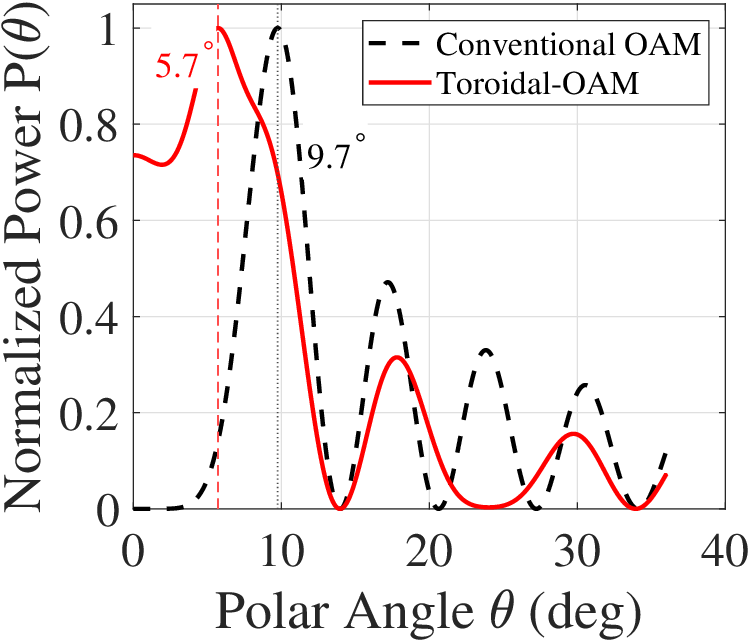}
\caption{Far-field angular power pattern $P(\theta)$ for $\ell=4$ at $\kap=1.0$, $\alp=0.7$. Toroidal-OAM (solid red) vs.\ conventional OAM (dashed black). Under strong coupling, the main lobe shift is dramatic, and the far-field pattern transforms from a pure annular lobe to a hybrid central-peak-plus-ring structure. Non-zero on-axis power for toroidal-OAM is visible near $\theta=0^\circ$.}
\label{fig:farfield}
\end{figure}

Fig.~\ref{fig:farfield} shows the far-field angular power pattern for $\ell=4$ as a representative example of the enhanced transition regime. The hybrid beam (solid red) exhibits a pronounced main-lobe shift toward smaller $\theta$ relative to conventional OAM (dashed black), with visible on-axis power filling from the $E_z$ pedestal. The corresponding patterns for $\ell=1$ and $\ell=8$ are consistent with the divergence values in Table~\ref{tab:divergence}: $\ell=1$ shows moderate inward shift with modest on-axis filling, while $\ell=8$ is fully center-peaked. The signed $E_z$ field, amplified by $\alpha=0.7$, provides constructive interference that pulls the effective main lobe inward even for high-order modes.

Under strong coupling, the three angular regimes evolve as follows: (i) \textbf{low-mode regime} ($\ell \leq 2$): the main lobe lies within or near $\theta_c$, yielding strong divergence suppression with moderate on-axis filling; (ii) \textbf{enhanced transition regime} ($\ell = 3$--$4$): the strongest relative divergence suppression for modes retaining a distinct OAM ring, enabled by the combined effect of filter attenuation and $E_z$ pedestal; (iii) \textbf{center-peaked regime} ($\ell \geq 5$): the OAM ring is fully suppressed and the beam becomes completely center-peaked, representing the most extreme form of divergence suppression where the longitudinal $E_z$ pedestal dominates the far-field power pattern entirely. The practical benefit extends meaningfully through $\ell=4$ where the divergence reduction reaches $41\%$.

\subsection{Communication Link Performance}

\begin{figure}[htbp]
\centering
\subfigure[Power capture efficiency vs.\ $z$ ($D_{\mathrm{RX}}=20\lambda$).]{
\begin{minipage}{0.45\linewidth}
\centering
\includegraphics[scale=0.32]{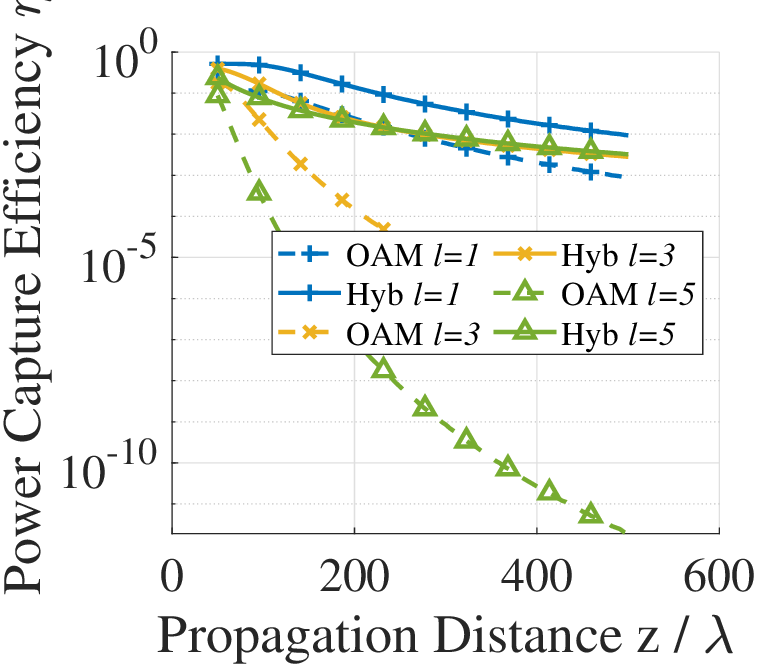}
\end{minipage}
}
\subfigure[SNR gain vs.\ $z$.]{
\begin{minipage}{0.45\linewidth}
\centering
\includegraphics[scale=0.32]{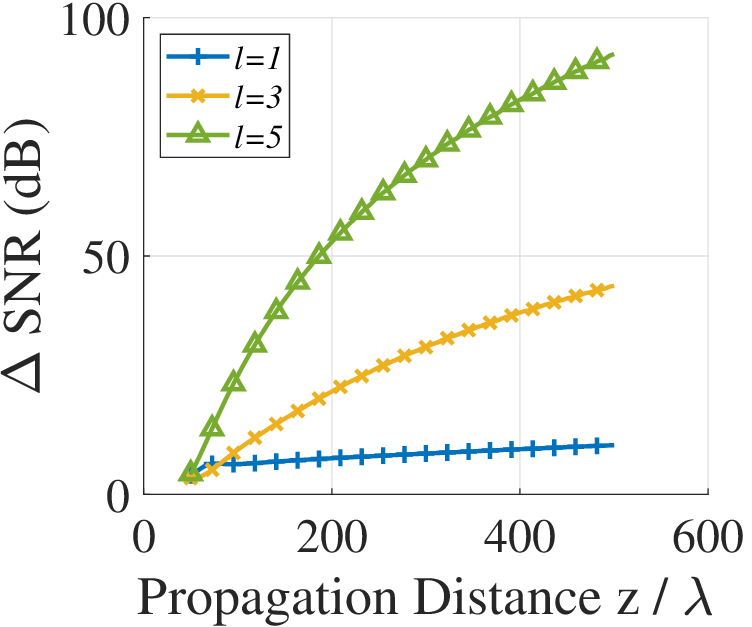}
\end{minipage}
}
\caption{Distance-dependent propagation analysis for $\ell=1$--$5$ ($\kap=1.0$, $\alp=0.7$, $D_{\mathrm{RX}}=20\lambda$). (a) Power capture efficiency vs.\ $z$ (log scale): toroidal-OAM preserves substantially higher $\eta$, with the advantage most pronounced at intermediate modes. (b) $\Delta$SNR grows with distance, positive for all $\ell=1$--$5$ throughout $z \geq 50\lambda$.}
\label{fig:distance}
\end{figure}

The divergence suppression quantified in Table~\ref{tab:divergence} directly reduces the beam spot radius $r_{\ell}(z) = z\tan(\theta_{\mathrm{div}}^{(\ell)})$ in the far field, which in turn increases the power capture efficiency $\eta_{\ell}$ at a finite-aperture receiver. Fig.~\ref{fig:distance} illustrates this distance-dependent behavior: toroidal-OAM preserves substantially higher $\eta$ than conventional OAM at all distances (most dramatically at $\ell=3$--$4$ where divergence reduction is strongest), and the resulting $\Delta$SNR grows with $z$, remaining positive for all $\ell=1$--$5$ throughout $z \geq 50\lambda$.

\begin{table}[htbp]
\centering
\caption{Power capture efficiency, SNR gain, and fixed-efficiency link-budget range equivalent ($D_{\mathrm{RX}} = 20\lambda$, $\kap=1.0$, $\alp=0.7$).}
\label{tab:link_performance}
\begin{tabular}{c c c c c c}
\toprule
$\ell$ & $z$ [$\lambda$] & $\eta_{\mathrm{OAM}}$ & $\eta_{\mathrm{Hyb}}$ & $\Delta\mathrm{SNR}$ [dB] & Range ext. \\
\midrule
1 & 100 & 0.109 & 0.474 & 6.4 & $+108\%$ \\
2 & 100 & 0.065 & 0.305 & 6.7 & $+117\%$ \\
3 & 100 & 0.017 & 0.151 & 9.4 & $+194\%$ \\
\midrule
1 & 300 & 0.0060 & 0.0435 & 8.6 & $+168\%$ \\
2 & 300 & $2.8\times 10^{-4}$ & $1.20\times 10^{-2}$ & 16.3 & $+554\%$ \\
3 & 300 & $6.7\times 10^{-6}$ & $8.26\times 10^{-3}$ & 30.9 & $+3405\%$ \\
\bottomrule
\end{tabular}
\end{table}

Table~\ref{tab:link_performance} translates beam-level improvements to communication metrics via the link budget framework of Section~\ref{sec:analysis}. Under $\alpha=0.7$, SNR gains of $6.4$--$30.9$~dB and fixed-efficiency link-budget range equivalent extension of $+108\%$ to $+3405\%$ (per Eq.~\eqref{eq:range_ext}, not a measured link range) are demonstrated, representing $44$--$64\%$ higher SNR gain and $63$--$112\%$ greater range extension compared with balanced coupling ($\alpha=0.5$). The compound benefit exceeds what the divergence reduction alone suggests, because $\eta_{\ell}$ is highly nonlinear in the divergence angle: a modest inward main-lobe shift produces a disproportionately large increase in integrated power within the receiver's narrow angular window. For a representative 3~GHz backhaul link ($P_{\mathrm{TX}} = 30$~dBm, $G = 15$~dBi, $B = 20$~MHz, $F = 6$~dB, $\ell=2$, $z = 300\lambda = 30$~m), conventional OAM achieves $\mathrm{SNR} \approx 10.0$~dB (QPSK capable), while toroidal-OAM provides $\mathrm{SNR} \approx 24.7$~dB (robust 64-QAM/256-QAM)---a two-step modulation-order upgrade enabled solely by the optimized toroidal beam modification.  Table~\ref{tab:link_performance} gives per-mode received-power SNR gains (Eq.~\eqref{eq:snr}). In the aligned link, azimuthally symmetric toroidal modifications induce inter-OAM-mode leakage below $10^{-9}$ (Section~\ref{sec:analysis}), leaving these gains essentially uneroded. The non-OAM pedestal contributes the power-fraction factor $S_{\ell}$ (Eq.~\eqref{eq:sinr}); the ideal/crosstalk-inclusive capacity gaps in Table~\ref{tab:capacity} are $8.14\%$, $5.30\%$, and $3.89\%$ at SNR $= 20$, $30$, and $40$~dB; per-mode modeled penalties $10\log_{10}(1/S_{\ell})$ are $0.53$, $1.07$, and $2.67$~dB for $\ell=1,2,3$.

\begin{table}[htbp]
\centering
\caption{Power imbalance and Jain fairness index for 3-mode OAM-MDM ($\ell = 1, 2, 3$, $\kap=1.0$, $\alp=0.7$).}
\label{tab:fairness}
\begin{tabular}{c c c c c}
\toprule
$z$ [$\lambda$] & OAM $\eta$ range & Hyb $\eta$ range & OAM JFI & Hyb JFI \\
\midrule
100  & $6.3\times$  & $3.1\times$  & 0.744 & 0.846 \\
300  & $9.0\times 10^2$ & $5.3\times$  & 0.365 & 0.645 \\
500  & $7.3\times 10^3$ & $3.3\times$  & 0.344 & 0.740 \\
1000 & $1.2\times 10^5$ & $2.0\times$  & 0.336 & 0.896 \\
\bottomrule
\end{tabular}
\end{table}

Table~\ref{tab:fairness} quantifies a second critical benefit of strong coupling: the toroidal-OAM hybrid dramatically reduces multi-mode power imbalance. At $z = 1000\lambda$, the capture efficiency ratio improves from $1.2\times 10^5$ to $2.0\times$ (a five-order-of-magnitude reduction) and the Jain fairness index improves from 0.34 (severely unbalanced) to 0.90 (near-ideal balance). At shorter distances, the toroidal-OAM JFI degrades more gracefully than conventional OAM, and at extreme distances the $E_z$ pedestal, which is $\ell$-independent, becomes the dominant received component, naturally equalizing the per-mode efficiency.

\subsection{Channel Capacity Results}

\begin{figure}[htbp]
\centering
\subfigure[Capacity vs SNR ($M=3$, $z=500\lambda$).]{
\begin{minipage}{0.45\linewidth}
\centering
\includegraphics[scale=0.28]{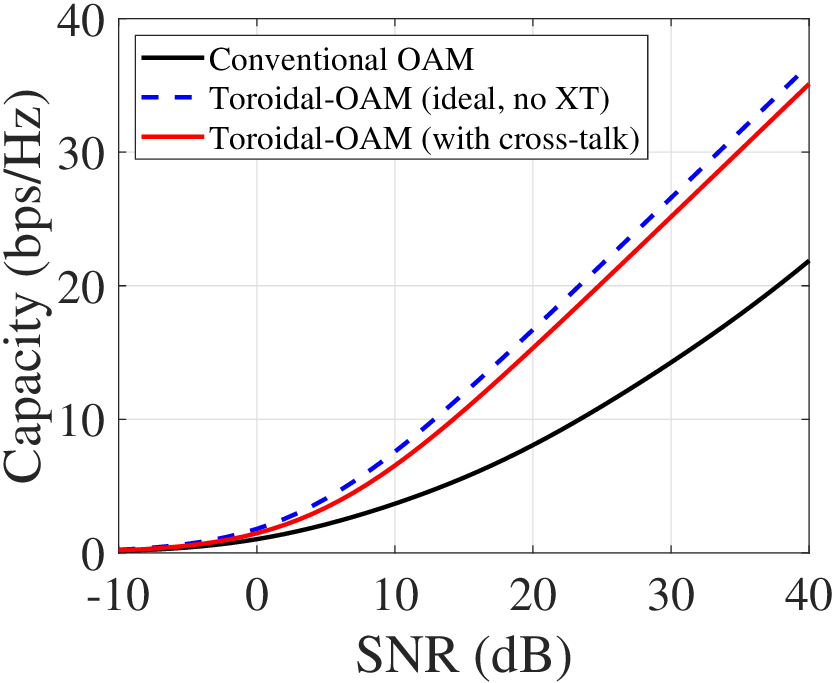}
\end{minipage}
}
\subfigure[Capacity vs mode count $M$ (SNR=20~dB).]{
\begin{minipage}{0.45\linewidth}
\centering
\includegraphics[scale=0.28]{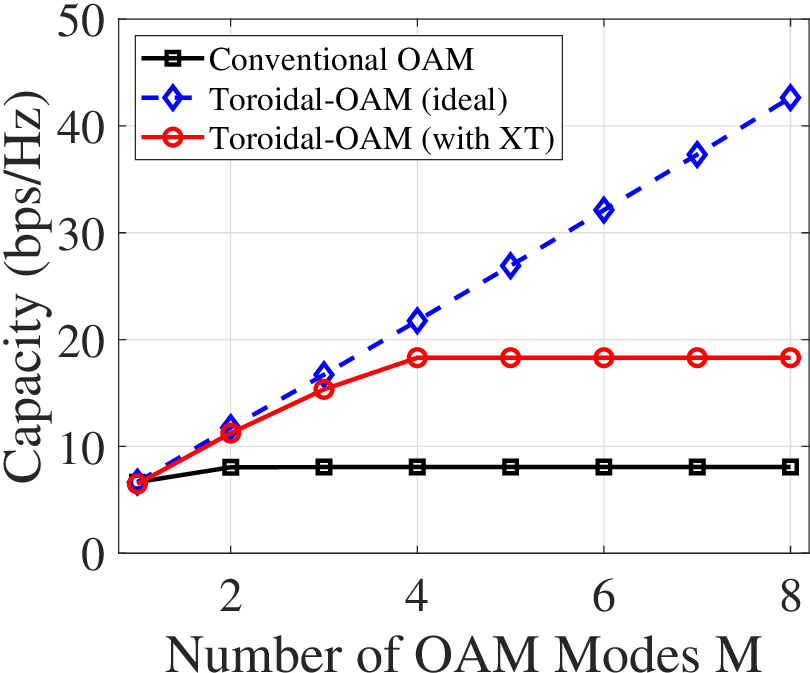}
\end{minipage}
}
\caption{Channel capacity analysis for $M=3$ OAM-MDM ($\ell=1,2,3$) at $z=500\lambda$, $\alp=0.7$, based on the channel model of Section~\ref{sec:capacity}. (a) Capacity vs.\ SNR: the toroidal-OAM ideal and cross-talk-inclusive curves remain close, and the ideal curve reaches a $107\%$ enhancement over conventional OAM at SNR=20~dB. (b) Capacity vs.\ mode count $M$ at SNR=20~dB: the ideal toroidal-OAM capacity sustains growth through $M=8$, the cross-talk-inclusive curve grows through $M=4$ and then saturates at $18.3$~bps/Hz, whereas conventional OAM capacity saturates at $8.1$~bps/Hz near $M=2$--$3$.}
\label{fig:capacity}
\end{figure}

\begin{table}[htbp]
\centering
\caption{Channel capacity comparison at $z=500\lambda$, $M=3$ ($\ell=1,2,3$), $\alp=0.7$. $C_{\mathrm{Hyb}}^{\mathrm{(XT)}}$ is close to $C_{\mathrm{Hyb}}^{\mathrm{(ideal)}}$ due to the azimuthal symmetry of the toroidal modifications.}
\label{tab:capacity}
\begin{tabular}{c c c c c}
\toprule
SNR (dB) & $C_{\mathrm{OAM}}$ & $C_{\mathrm{Hyb}}^{\mathrm{(ideal)}}$ & $C_{\mathrm{Hyb}}^{\mathrm{(XT)}}$ & $\Delta C$ \\
\midrule
10 & 3.68 & 7.57 & 6.54 & 3.90 \\
20 & 8.07 & 16.71 & 15.35 & 8.64 \\
30 & 14.26 & 26.58 & 25.17 & 12.33 \\
40 & 21.88 & 36.54 & 35.12 & 14.66 \\
\bottomrule
\end{tabular}
\end{table}

Fig.~\ref{fig:capacity} and Table~\ref{tab:capacity} present the channel capacity analysis under $\alpha=0.7$ at $z=500\lambda$. At SNR=20~dB, the ergodic capacity increases from 8.07~bps/Hz (conventional OAM, $M=3$) to 16.71~bps/Hz (toroidal-OAM ideal), a $107\%$ enhancement driven by the power capture efficiency improvement. The capacity gain grows with SNR, demonstrating that the hybrid beam advantage compounds at higher SNR and longer distances.  In Table~\ref{tab:capacity}, the $\Delta C$ column reports the ideal capacity gain $C_{\mathrm{Hyb}}^{\mathrm{(ideal)}} - C_{\mathrm{OAM}}$ (e.g., $16.71 - 8.07 = 8.64$~bps/Hz at SNR$=20$~dB). The crosstalk-inclusive gain is $C_{\mathrm{Hyb}}^{\mathrm{(XT)}} - C_{\mathrm{OAM}}$ (e.g., $15.35 - 8.07 = 7.28$~bps/Hz at SNR$=20$~dB).

 We emphasize that the capacity gain of the hybrid beam originates from the beam-level power capture improvement rather than from an increase in the number of spatial degrees of freedom: both schemes multiplex the same $M$ modes, and the hybrid raises the modeled per-mode capture metric. Table~\ref{tab:capacity} applies the per-scheme peak normalization stated in Section~\ref{sec:capacity}. Under a common absolute normalization, the ideal capacity gain is $1.59$~bps/Hz at SNR$=20$~dB and $12.98$~bps/Hz at SNR$=40$~dB, consistent with the interpretation that the advantage is a power capture gain that grows with the operating SNR.

The cross-talk-inclusive capacity $C_{\mathrm{Hyb}}^{\mathrm{(XT)}}$ depends on the mode purity spectrum $\{S_n\}$ (Eq.~\ref{eq:mode_spectrum}).  At $\alpha=0.7$, the intended-mode power fraction at the ring is $S_{\ell} = 0.885$ for $\ell=1$, $0.781$ for $\ell=2$, and $0.541$ for $\ell=3$; for $\ell \geq 4$ the beam becomes pedestal-dominated. At $\alpha=0.5$, $S_{\ell} = 0.940$, $0.896$, and $0.811$. The toroidal modifications are azimuthally symmetric and do not mix OAM modes: the computed inter-OAM-mode leakage is below $10^{-9}$, and the deficit of $S_{\ell}$ from unity is the power carried by the azimuthally symmetric non-OAM pedestal. Consequently, $C_{\mathrm{Hyb}}^{\mathrm{(XT)}}$ differs from $C_{\mathrm{Hyb}}^{\mathrm{(ideal)}}$ by the per-mode power-fraction loss quantified in Table~\ref{tab:capacity}, and the XT and ideal curves in Fig.~\ref{fig:capacity}(a) remain close.

Fig.~\ref{fig:capacity}(b) examines capacity scaling with mode count $M$ at a fixed SNR of 20~dB. Under $\alpha=0.7$, the ideal toroidal-OAM capacity sustains growth through $M=8$ ($42.7$~bps/Hz at $M=8$), while the cross-talk-inclusive capacity grows through $M=4$ and then saturates at $18.3$~bps/Hz, because the pedestal-dominated modes $\ell \geq 4$ carry little intended-mode power; conventional OAM capacity saturates at $8.1$~bps/Hz near $M=2$--$3$, since the diminishing per-mode received power for $\ell \geq 3$ adds negligible rate. At $M=8$ the cross-talk-inclusive advantage is $18.3$ versus $8.1$~bps/Hz ($2.3\times$), and the ideal advantage is $42.7$ versus $8.1$~bps/Hz ($5.3\times$). The toroidal-OAM capacity advantage at large $M$ is a direct consequence of the mode equalization effect (Table~\ref{tab:fairness}): higher-order modes remain viable communication channels under strong coupling, whereas they effectively vanish for conventional OAM.

A systematic parameter sweep across $R/\lambda \in [3,6]$, $\kappa \in [0.5, 2.5]$, $\alpha \in [0.1, 0.9]$, and $D_{\mathrm{RX}}/\lambda \in [10, 50]$ identifies three coupling regimes:
\begin{itemize}
  \item \textbf{Weak coupling ($\alpha \leq 0.3$)}: Near-pure OAM operation with minimal beam modification. Divergence ratio $> 0.95$ for most $\ell$, $R_{\mathrm{ring}} > 50$. Suitable for high-fidelity multi-mode applications.
  \item \textbf{Balanced regime ($\alpha \approx 0.5$, $\kappa \approx 1.0$)}: Moderate divergence reduction ($6$--$22\%$) while maintaining a clearly ring-shaped profile ($R_{\mathrm{ring}} > 10$).
  \item \textbf{Strong coupling ($\alpha = 0.7$, $\kappa \approx 1.0$)}: \textbf{Recommended optimized configuration.} Maximizes all key metrics simultaneously (see Table~\ref{tab:alpha_comparison}). Divergence reduction reaches $41\%$ for $\ell=4$, modes $\ell \geq 5$ become fully center-peaked, and capacity gain reaches $107\%$ at SNR=20~dB. Recommended for single-mode and few-mode ($M \leq 3$) links.
\end{itemize}

\begin{table}[htbp]
\centering
\caption{ Quantitative comparison: $\alp=0.5$ (balanced coupling) vs.\ $\alp=0.7$ (strong coupling, recommended). All values at $\kap=1.0$, $R=5\lambda$, $f=3$~GHz unless otherwise noted.}
\label{tab:alpha_comparison}
\resizebox{\linewidth}{!}{
    \begin{tabular}{l c c}
    \toprule
    Metric & $\alp = 0.5$ & $\alp = 0.7$ \\
    \midrule
    Mean divergence ratio $\langle\thetadiv^{\mathrm{(Hyb)}}/\thetadiv^{\mathrm{(OAM)}}\rangle$ & 0.665 & 0.364 \\
    Max.\ divergence reduction & $22\%$ ($\ell=3$) & $41.4\%$ ($\ell=4$) \\
    On-axis field $P(0)/P_{\max}$ at $\ell=1$ & $<1\%$ & $8.0\%$ \\
    On-axis field at $\ell=5$ & $11\%$ & $100\%$ \\
    $R_{\mathrm{ring}}$ at $\ell=1$ & $52.9$ & $12.44$ \\
    $R_{\mathrm{ring}}$ at $\ell=8$ & $9.3$ & $1.00$ \\
    Azimuthal intended-mode power fraction $S_{\ell}$ (range) & $0.811$--$0.940$ & $0.541$--$0.885$ \\
    $\ell=1$, $z=100\lambda$: $\Delta\mathrm{SNR}$ & $3.8$~dB & $6.4$~dB \\
    $\ell=3$, $z=300\lambda$: $\Delta\mathrm{SNR}$ & $17.5$~dB & $30.9$~dB \\
    JFI at $z=100\lambda$ & $0.823$ & $0.846$ \\
    JFI at $z=1000\lambda$ & $0.60$ & $0.896$ \\
    $C_{\mathrm{Hyb}}^{\mathrm{(ideal)}}$ at SNR=20~dB, $M=3$ & 8.07~bps/Hz & 16.71~bps/Hz \\
    Capacity gain $\Delta C$ at SNR=20~dB & $-$ & $8.64$~bps/Hz ($107\%$) \\
    \bottomrule
    \end{tabular}
}
\end{table}

Table~\ref{tab:alpha_comparison} summarizes the quantitative differences between the balanced and strong coupling configurations. The \textbf{fundamental tradeoff} is beam morphology vs.\ communication performance: larger $\alpha$ increases the on-axis field and inward ring shift while reducing $R_{\mathrm{ring}}$, but azimuthal mode orthogonality is preserved because $T(\theta)$ and $E_z(\theta)$ are azimuthally symmetric. The primary cost is morphological: at $\alpha=0.7$, modes $\ell \geq 5$ are fully center-peaked, while for $\ell=1$--$3$, $R_{\mathrm{ring}} > 2.8$ and the ring structure remains clearly dominant. The sweep confirms that $\alpha=0.7$ represents a favorable operating point where all key communication metrics improve simultaneously.

\subsection{Misalignment Robustness}
\label{sec:misalignment}

Practical OAM links are sensitive to lateral displacement and pointing errors because reception requires matching the helical phase across the receive aperture. We evaluate robustness through the matched-mode capture efficiency, the power delivered to the intended mode by conjugate azimuthal demultiplexing,
\begin{equation*}
\eta_{\ell}^{\mathrm{(mode)}} = \frac{\bigl|\int_{A_{\mathrm{RX}}} E(x,y)\, e^{-i\ell\phi'}\, \mathrm{d}A\bigr|^2}{A_{\mathrm{RX}}\, P_{\mathrm{sim}}},
\end{equation*}
where $A_{\mathrm{RX}}$ is the receive aperture, $P_{\mathrm{sim}}$ is the total power integrated over the finite transverse simulation window, and $\phi'$ is measured about the receive axis. Lateral displacement translates the aperture while keeping its axis parallel to the beam axis. Pointing error tilts the demultiplexing reference axis by $\psi$ relative to the beam axis on a fixed aperture, a small-angle approximation; both cases use the scalar far-field model of Eq.~\eqref{eq:field_2d}. The conjugate-weighted disk represents the fixed-DFT demultiplexer of Section~\ref{sec:capacity} in this scalar model. Its matched-mode efficiency equals power capture efficiency only for a constant-amplitude mode profile across the aperture, while the normalized sensitivity curves are independent of $P_{\mathrm{sim}}$.

\begin{figure}[htbp]
\centering
\subfigure[Matched-mode capture vs.\ lateral displacement.]{
\begin{minipage}{0.45\linewidth}
\centering
\includegraphics[scale=0.28]{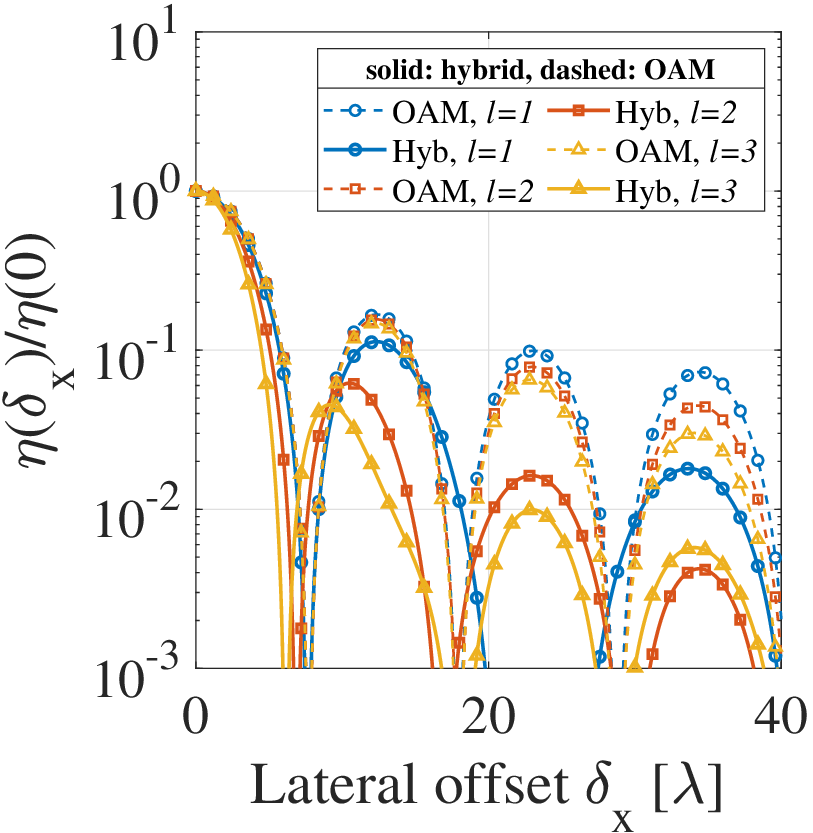}
\end{minipage}
}
\subfigure[Matched-mode capture vs.\ pointing error.]{
\begin{minipage}{0.45\linewidth}
\centering
\includegraphics[scale=0.28]{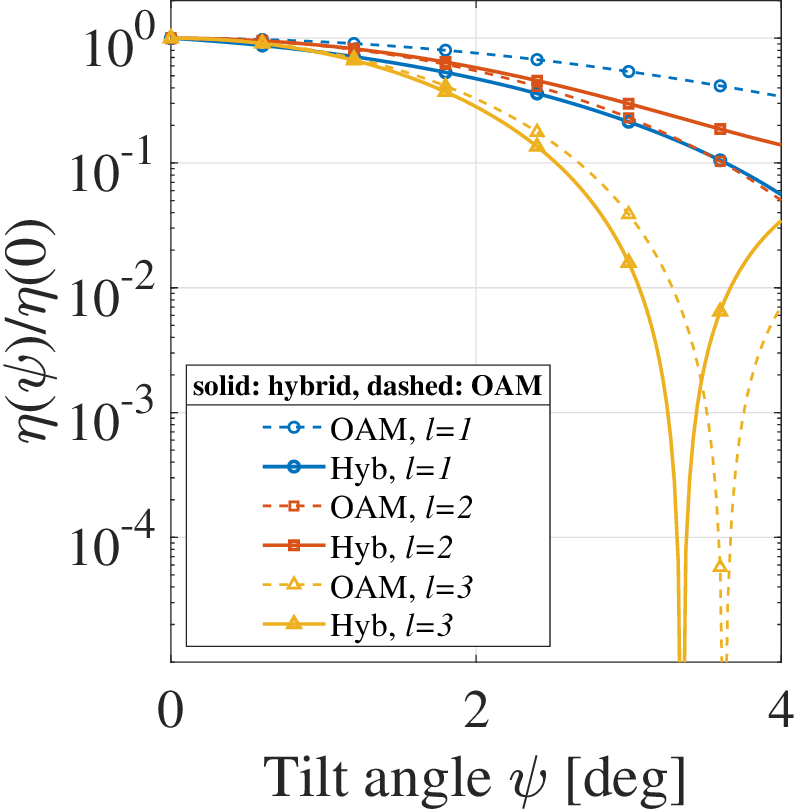}
\end{minipage}
}\\
\subfigure[Hybrid capture normalized by the aligned conventional-OAM value.]{
\begin{minipage}{1\linewidth}
\centering
\includegraphics[scale=0.28]{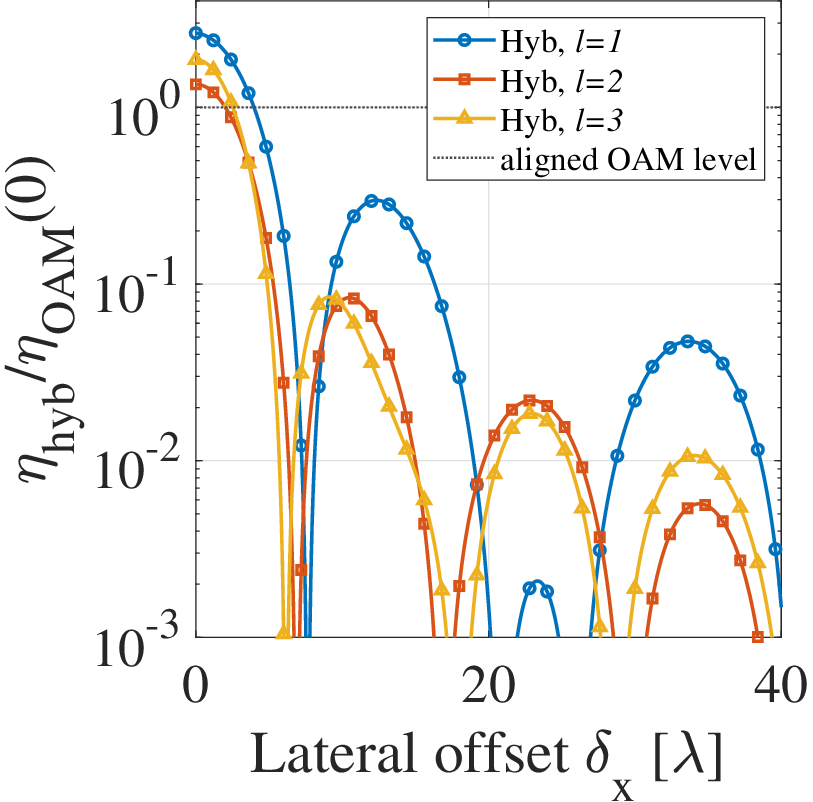}
\end{minipage}
}
\caption{ Misalignment robustness of the toroidal-OAM hybrid beam at $z=100\lambda$, $D_{\mathrm{RX}}=20\lambda$, $\kappa=1.0$, $\alpha=0.7$. (a) Normalized matched-mode capture versus lateral displacement of the receive aperture; dashed curves: conventional OAM, solid curves: hybrid beam, for $\ell=1,2,3$. (b) Normalized matched-mode capture versus pointing error of the receive axis. (c) Hybrid matched-mode capture normalized by the aligned conventional-OAM value.}
\label{fig:misalignment}
\end{figure}

Fig.~\ref{fig:misalignment} plots normalized matched-mode capture versus (a) lateral displacement and (b) pointing error for $\ell=1,2,3$ at $z=100\lambda$ with $D_{\mathrm{RX}}=20\lambda$; panel~(c) normalizes hybrid capture by aligned conventional-OAM capture. The sensitivities are comparable: the hybrid $-3$~dB lateral tolerance is $0.13$--$0.17\,D_{\mathrm{RX}}$, versus $0.18\,D_{\mathrm{RX}}$ for conventional OAM, and the $-3$~dB tilt tolerance is $1.54$--$2.26^{\circ}$ versus $1.61$--$3.19^{\circ}$ (Table~\ref{tab:misalignment}). The aligned hybrid gains of $4.2$, $1.3$, and $2.7$~dB for $\ell=1,2,3$ provide displacement margin: the hybrid matches aligned conventional-OAM capture up to offsets of approximately $0.10$--$0.20$ receiver diameters ($2$--$4$ wavelengths). Beyond these offsets the advantage is lost; at one receiver diameter, relative gains are $-13.8$, $-4.1$, and $-6.4$~dB. These results use a fixed-DFT receiver without alignment correction. The aligned capture advantage supporting Section~\ref{sec:capacity} is therefore subject to mode-dependent alignment tolerances of comparable scale for both schemes; adaptive-alignment performance is not evaluated.

\begin{table*}[htbp]
\centering
\caption{ Misalignment tolerance comparison ($z=100\lambda$, $D_{\mathrm{RX}}=20\lambda$, $\kappa=1.0$, $\alpha=0.7$): $-3$~dB lateral offset and $-3$~dB pointing error in matched-mode capture, and the matched-mode capture gain of the hybrid over conventional OAM at zero and one receiver diameter of displacement.}
\label{tab:misalignment}
\setlength{\tabcolsep}{4pt}
\begin{tabular}{c c c c c c c}
\toprule
$\ell$ & Hyb offset [$D_{\mathrm{RX}}$] & OAM offset [$D_{\mathrm{RX}}$] & Hyb tilt [$^\circ$] & OAM tilt [$^\circ$] & $\Delta$SNR $(\delta_x=0)$ [dB] & $\Delta$SNR $(\delta_x=D_{\mathrm{RX}})$ [dB] \\
\midrule
1 & 0.17 & 0.18 & 1.91 & 3.19 & 4.2 & $-13.8$ \\
2 & 0.15 & 0.18 & 2.26 & 2.14 & 1.3 & $-4.1$ \\
3 & 0.13 & 0.18 & 1.54 & 1.61 & 2.7 & $-6.4$ \\
\bottomrule
\end{tabular}
\end{table*}



\section{Full-Wave Simulation Verification}
\label{sec:hfss}

\begin{figure}[htbp]
\centering
\subfigure[HFSS model: coaxial horn (center) and UCA monopole ring.]{
\includegraphics[scale=0.09]{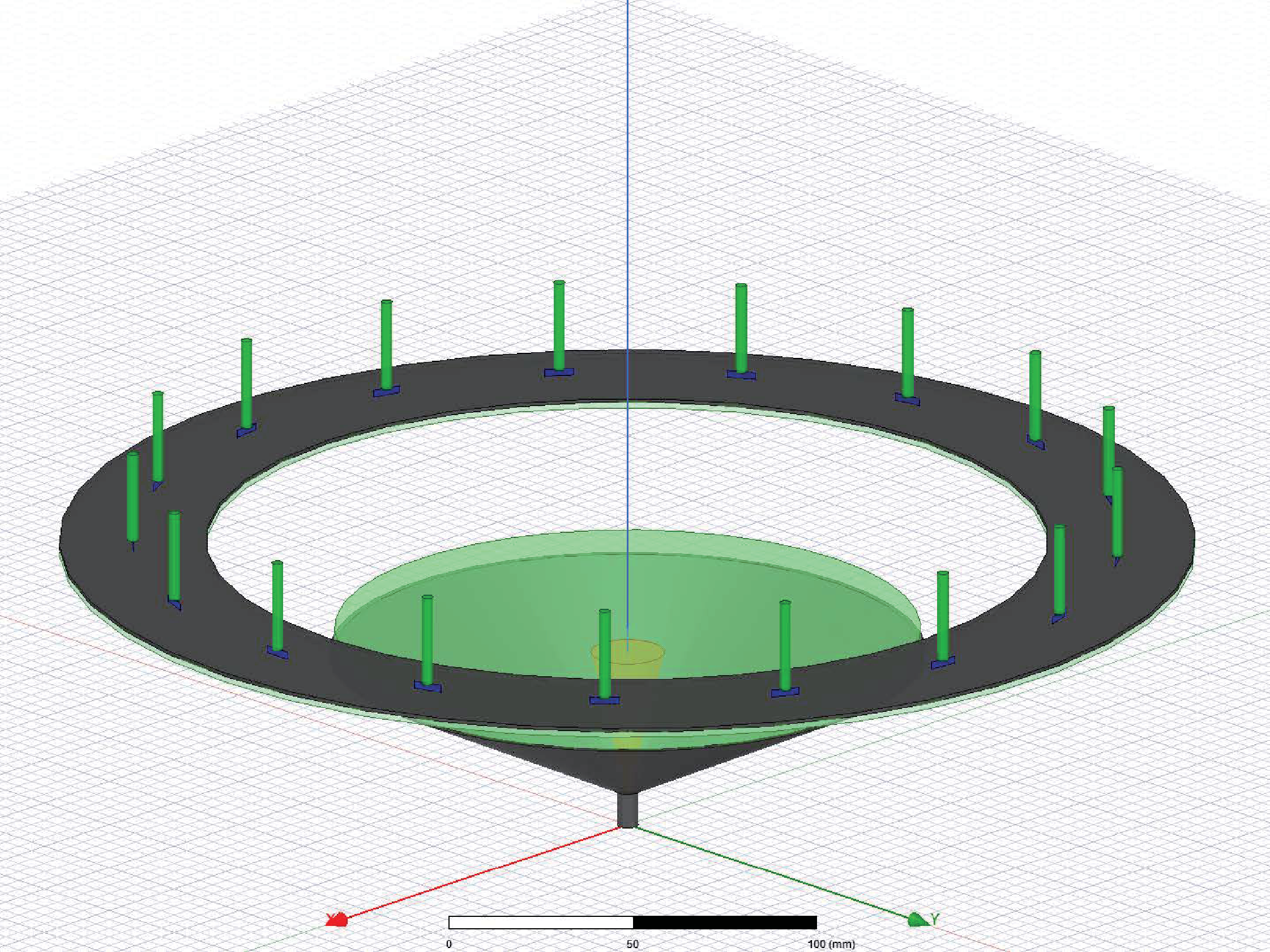}
}
\subfigure[Detail of the coaxial horn feed and flare structure.]{
\includegraphics[scale=0.15]{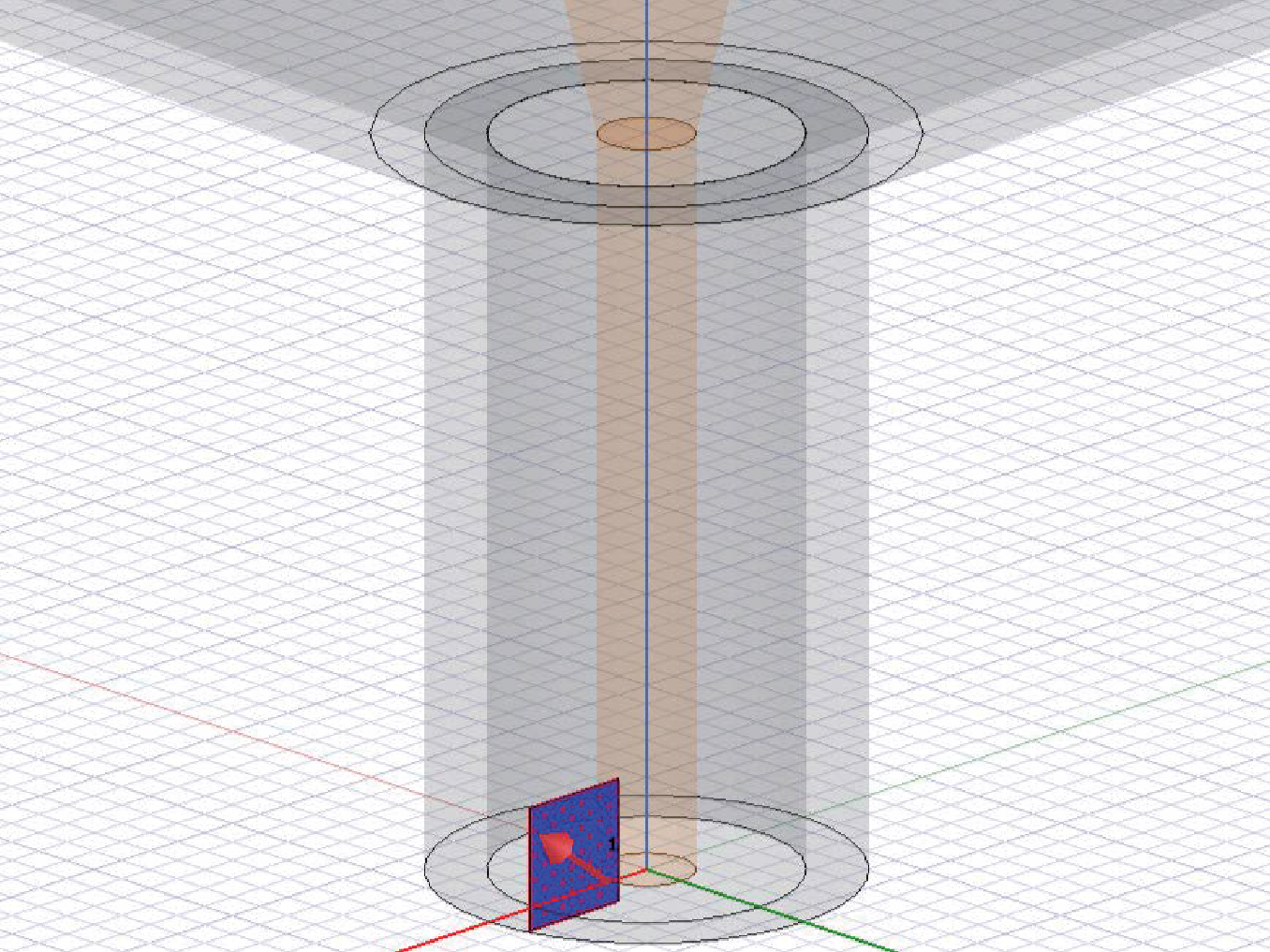}
}
\caption{ANSYS HFSS full-wave model of the toroidal-OAM hybrid beam launcher. (a) Overall geometry showing the central coaxial horn and the surrounding 17-element UCA ring. (b) Cross-sectional detail of the horn feed and conical flare.}
\label{fig:hfss_model}
\end{figure}
The analytical model of Section~\ref{sec:analysis} is a phenomenological far-field description. To verify that the underlying physical concept (a coaxial horn co-located with an OAM-UCA on a shared aperture) produces the predicted beam characteristics in a full-wave electromagnetic environment, a three-dimensional model of the hybrid launcher was constructed and solved in ANSYS HFSS 2020 R2. The model comprises a wide-angle short coaxial horn antenna (inner/outer conductors expanding from 0.635/2.85~mm to 10/80~mm over 40~mm, half-flare angle $\approx 63^\circ$) co-located with a 17-element UCA of vertical monopoles at radius $R = 17\lambda/(4\pi) \approx 1.35\lambda$. The horn port and the 17 UCA ports are excited with the power split $P_{\mathrm{horn}} = 0.7$~W and $P_{\mathrm{UCA}} = 0.3/17$~W per element (a $70\%$ horn power share, $r_p = 7/3$), which corresponds through Eq.~\eqref{eq:alpha_power} to an amplitude coupling coefficient $\alpha = \sqrt{r_p}/(1+\sqrt{r_p}) \approx 0.60$, and the UCA ports carry a progressive OAM phase $\phi_n = 2\pi\ell n/17$. Radiation boundaries are assigned to a cylindrical air enclosure, and the far-field is computed on an infinite sphere with $1^\circ$ angular resolution at 3~GHz with a discrete sweep from 2.9~GHz to 3.1~GHz.  Standard fabrication suffices for the wide-angle conical metal coaxial horn (machining or electroforming) and monopole UCA (printed or mounted). A single-layer metasurface could synthesize the filtered profile; narrow-divergence multi-OAM metasurfaces have been studied through compact vortex-structure measurements~\cite{lin2021dualmeta} and full-wave designs~\cite{li2021metasurface}. Planar metasurface realization of the complete hybrid, including its longitudinal component, remains proposed; electronic reconfiguration remains prospective.

Two sets of simulations were performed for each OAM mode $\ell = 1, 2, 3$, with total input power normalized to 1~W: (i) the full hybrid launcher with the horn carrying $70\%$ of the input power ($0.7$~W) and the remainder equally divided among the 17 UCA ports, following Eq.~\eqref{eq:alpha_power}, and (ii) a UCA-only baseline with the horn port de-energized and the full 1~W equally divided among the 17 UCA ports, preserving the OAM progressive phase. The far-field $|E(\theta)|$ was exported along the $\phi = 0^\circ$ cut.

\begin{table}[htbp]
\centering
\caption{Full-wave simulation: divergence angle and on-axis intensity comparison between the hybrid launcher and the UCA-only baseline.}
\label{tab:hfss_divergence}
\begin{tabular}{c c c c c c}
\toprule
$\ell$ & $\theta_{\mathrm{div}}^{\mathrm{(UCA)}}$ & $\theta_{\mathrm{div}}^{\mathrm{(Hyb)}}$ & $\Delta\theta$ & $P_{\mathrm{Hyb}}(0)/P_{\max}$ & $P_{\mathrm{UCA}}(0)/P_{\max}$ \\
\midrule
1 & $41.0^\circ$ & $36.0^\circ$ & $-5.0^\circ$ & 0.071 & 0.458 \\
2 & $54.0^\circ$ & $24.0^\circ$ & $-30.0^\circ$ & $<0.001$ & 0.002 \\
3 & $69.0^\circ$ & $31.0^\circ$ & $-38.0^\circ$ & $<0.001$ & $<0.001$ \\
\bottomrule
\end{tabular}
\end{table}
Table~\ref{tab:hfss_divergence} summarizes the divergence angles extracted from the far-field power patterns. Across all three modes, the hybrid launcher achieves a measurable reduction in the main-lobe angular position compared with the UCA-only case. For $\ell = 1$, the reduction is $5.0^\circ$ ($12\%$); for $\ell = 2$ and $\ell = 3$, the reductions are $30.0^\circ$ and $38.0^\circ$, respectively. The on-axis intensity $P(0)/P_{\max}$ of the hybrid launcher is $7.1\%$ at $\ell = 1$, demonstrating the central null filling contributed by the horn's longitudinal field component. The UCA-only baseline exhibits a higher on-axis value at $\ell=1$ ($45.8\%$), which is attributed to near-field coupling between the closely spaced monopole elements at the relatively small UCA radius of $1.35\lambda$, producing a non-ideal OAM phase front that partially fills the central null.

Table~\ref{tab:hfss_validation} reports two metrics extracted from HFSS $\varphi=0$ far-field cuts, treated as rotational averages under axisymmetry and normalized over the forward hemisphere. The resulting capture gains are forward-power concentration estimates obtained by integrating over the receive cone of a $20\lambda$ receiver at $z=100\lambda$. Relative to the UCA-only baseline, the hybrid gains $+16.8$ and $+14.5$~dB for $\ell=2,3$, whose main lobes move from $54^{\circ}$ and $69^{\circ}$ to $24^{\circ}$ and $31^{\circ}$. For $\ell=1$, the estimate is $-6.6$~dB because the simulated baseline has strong near-axis content: its $45.8\%$ on-axis level does not correspond to a genuine OAM beam, limiting comparison with ideal OAM. Although Table~\ref{tab:link_performance} predicts positive mode-one gain under its analytical assumptions, these HFSS estimates establish concentration improvements for modes two and three. The hybrid $-3$~dB main-lobe beamwidth is $23$--$30^{\circ}$ across $\ell=1$--$3$, substantially narrower than the baseline for $\ell=3$. These metrics support the modeled divergence suppression and axial power concentration, connecting the full-wave patterns to the capture analysis of Section~\ref{sec:analysis}, subject to the stated normalization and baseline limitations.

\begin{table}[htbp]
\centering
\caption{ Additional full-wave validation metrics extracted from the HFSS far-field patterns ($D_{\mathrm{RX}}=20\lambda$, $z=100\lambda$): capture gain of the hybrid launcher over the UCA-only baseline and $-3$~dB main-lobe beamwidth.}
\label{tab:hfss_validation}
\resizebox{\linewidth}{!}{
\begin{tabular}{c c c c c}
\toprule
$\ell$ & $\Delta\mathrm{SNR}$ [dB] & Hyb $-3$~dB beamwidth [$^\circ$] & UCA $-3$~dB beamwidth [$^\circ$] & $\theta_{\mathrm{div}}^{\mathrm{(Hyb)}}$ [$^\circ$] \\
\midrule
1 & $-6.6$ & 30 & 16 & 36 \\
2 & $+16.8$ & 23 & 21 & 24 \\
3 & $+14.5$ & 30 & 63 & 31 \\
\bottomrule
\end{tabular}
}
\end{table}

\begin{figure}[htbp]
\centering
\subfigure[$\ell=1$, hybrid launcher.]{\includegraphics[scale=0.1]{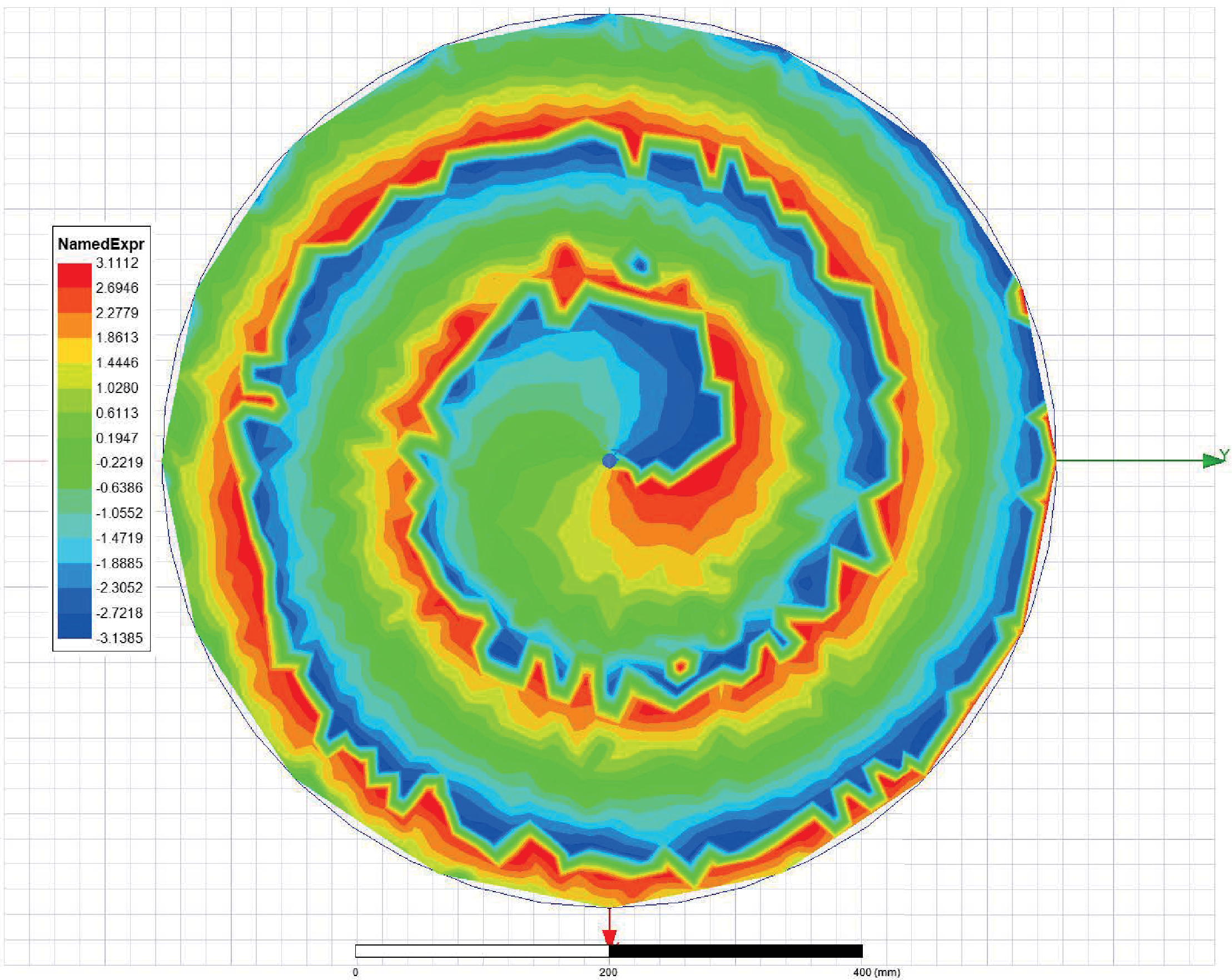}}
\subfigure[$\ell=2$, hybrid launcher.]{\includegraphics[scale=0.1]{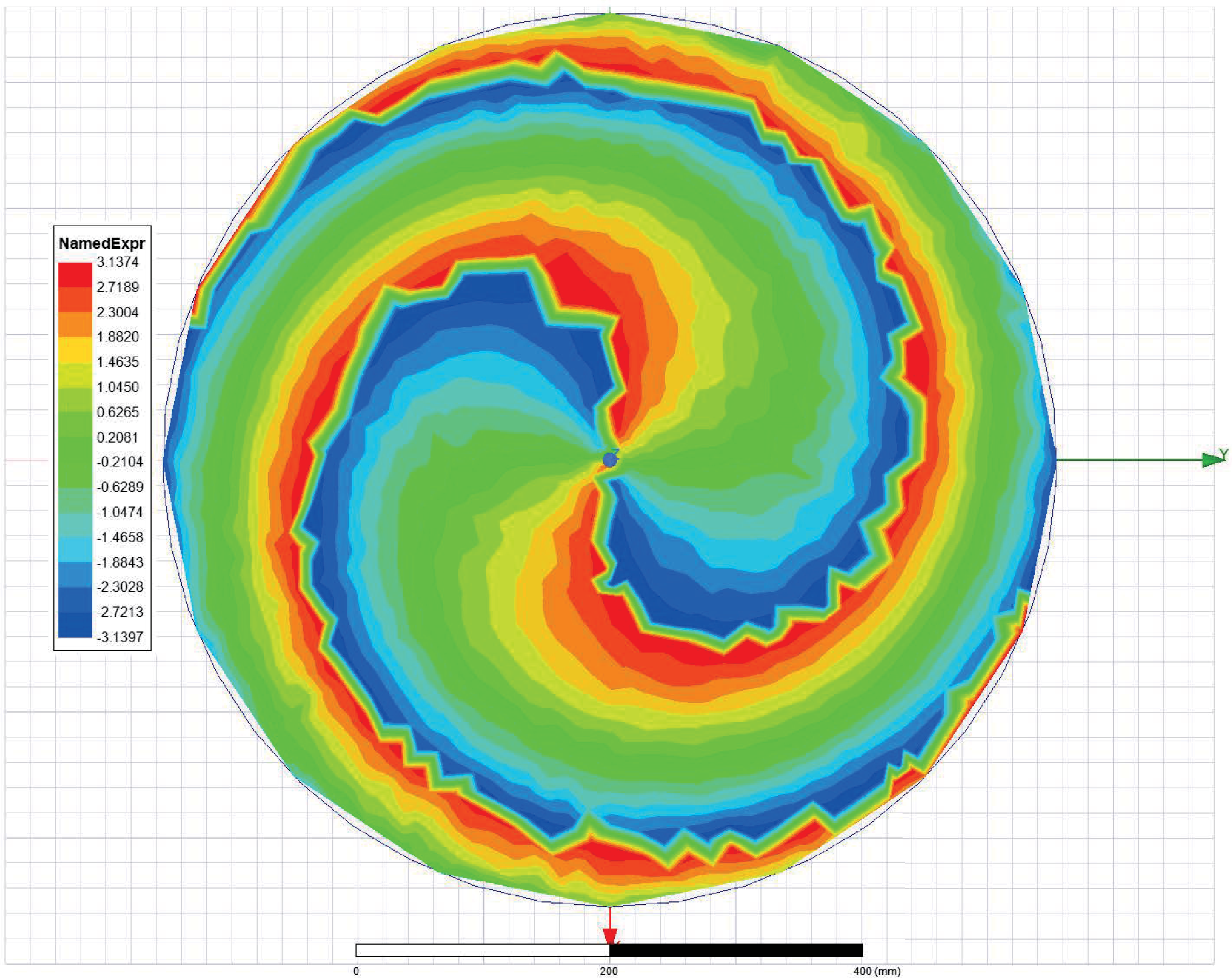}}
\subfigure[$\ell=3$, hybrid launcher.]{\includegraphics[scale=0.1]{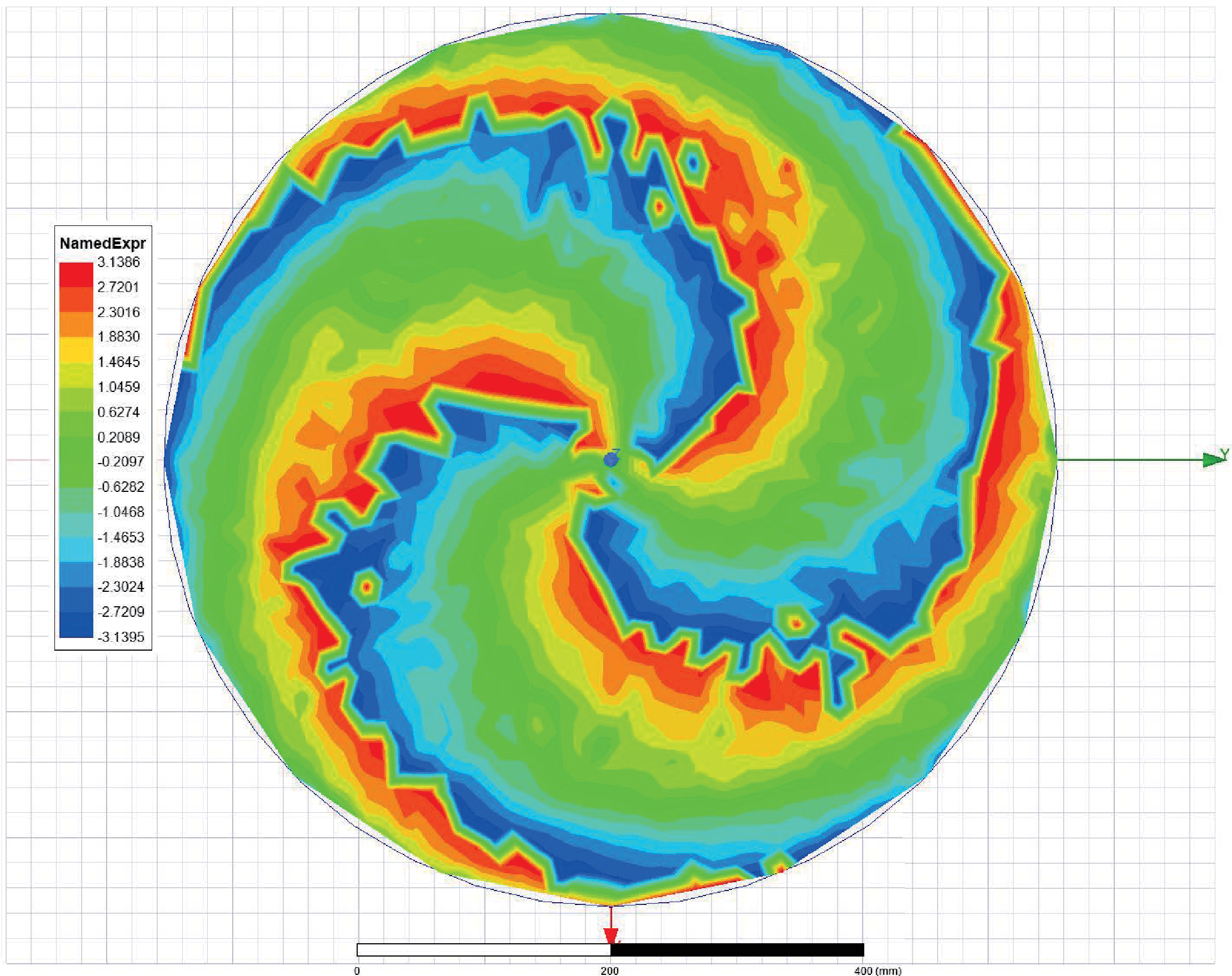}}
\caption{Phase distribution of the azimuthal electric field component $E_\phi$ in the transverse plane at $z = 25\lambda$, computed from HFSS field data via $E_\phi = E_y\cos\phi - E_x\sin\phi$. (a) $\ell=1$, (b) $\ell=2$, (c) $\ell=3$ hybrid launcher. The $\ell$-arm spiral structure is clearly visible in all three cases, confirming that the addition of the coaxial horn does not disrupt the OAM vortex phase topology.}
\label{fig:hfss_phase}
\end{figure}
To verify that the coaxial horn does not disrupt the OAM vortex phase structure, the azimuthal electric field component $E_\phi = E_y\cos\phi - E_x\sin\phi$ was computed from the HFSS near-field data on a transverse observation plane at $z = 25\lambda$ as shown in Fig.~\ref{fig:hfss_phase}. The phase of $E_\phi$ exhibits the characteristic $\ell$-arm spiral pattern at the UCA ring radius for the hybrid launcher, directly confirming that the OAM azimuthal phase $e^{i\ell\phi}$ is preserved when the horn is energized. This result is consistent with the analytical model's prediction that the toroidal modifications---the super-Gaussian angular filter $T(\theta)$ and the longitudinal $J_0$ pedestal---are azimuthally symmetric and therefore do not mix OAM modes in the azimuthal domain.

\begin{figure}[htbp]
\centering
\subfigure[$\ell=1$.]{\includegraphics[scale=0.26]{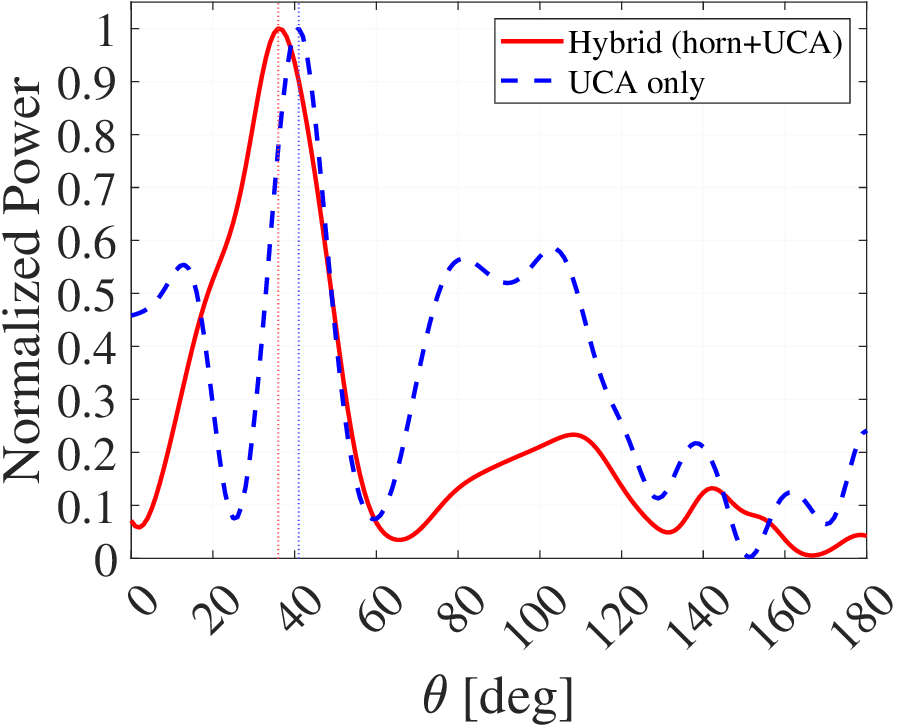}}
\subfigure[$\ell=2$.]{\includegraphics[scale=0.26]{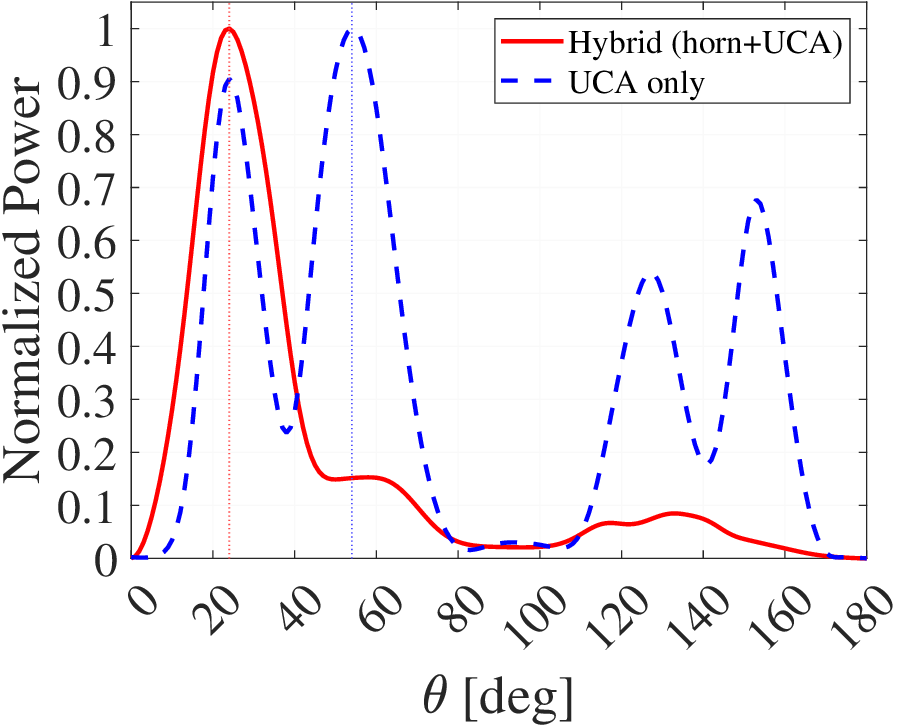}}
\subfigure[$\ell=3$.]{\includegraphics[scale=0.26]{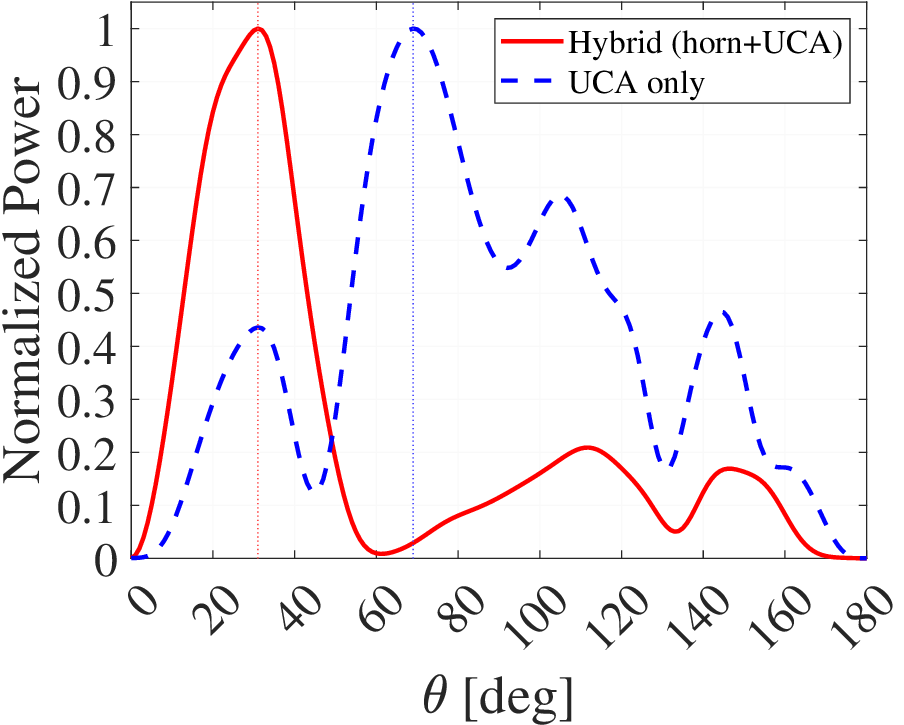}}
\caption{Normalized far-field power patterns $|E(\theta)|^2$ at $\phi = 0^\circ$ for the hybrid launcher (solid red) and the UCA-only baseline (dashed blue), exported from HFSS for $\ell = 1, 2, 3$. Vertical dashed lines mark the main-lobe divergence angle for each configuration.}
\label{fig:hfss_farfield}
\end{figure}
Fig.~\ref{fig:hfss_farfield} presents the normalized far-field power patterns for the three simulated OAM modes. In each subfigure, the hybrid launcher (solid red) exhibits a main lobe visibly shifted toward smaller polar angles relative to the UCA-only baseline (dashed blue), with the angular separation growing as $\ell$ increases. The vertical dotted lines provide a direct visual gauge of this divergence reduction, quantified in Table~\ref{tab:hfss_divergence}. At $\ell=1$, the hybrid pattern displays modest on-axis elevation relative to its own peak, consistent with partial central null filling; the UCA-only baseline exhibits a higher on-axis value ($45.8\%$ of peak, Table~\ref{tab:hfss_divergence}) attributed to near-field mutual coupling at the small array radius of $1.35\lambda$. For $\ell=2$ and $\ell=3$, on-axis intensities of both configurations are below $1\%$ of their respective peaks. The far-field patterns qualitatively support the divergence suppression mechanism described by the analytical model.

The divergence reduction observed in HFSS is qualitatively consistent with the analytical model's prediction: the addition of the coaxial horn modifies the far-field angular power distribution in a manner that concentrates radiated power toward smaller polar angles. The quantitative differences between the HFSS results and the analytical predictions of Section~\ref{sec:results} arise primarily from the smaller UCA radius ($1.35\lambda$ vs.\ $5\lambda$) and the wider horn aperture angle in the simulated geometry, which together shift the operating point to a regime of weaker toroidal confinement : for horn outer conductor radius $r_T = 80$~mm and UCA radius $R = 1.35\lambda \approx 135$~mm, Eq.~\eqref{eq:kappa_geometry} gives $\kappa = r_T/(2R) \approx 0.3$ (analytical reference configuration: $\kappa = 1.0$). These differences do not invalidate the analytical model, but rather illustrate that the model's three parameters ($\kappa$, $\alpha$, $\beta$) can be calibrated to describe a range of physical realizations of the hybrid launcher concept.

The full-wave comparisons follow Section~\ref{sec:results}: Table~\ref{tab:hfss_divergence} evaluates the divergence comparison of Section~V-C, and Table~\ref{tab:hfss_validation} evaluates the capture comparison of Section~V-F. Divergence decreases by $12\%$, $56\%$, and $55\%$ for $\ell=1,2,3$, reproducing the stronger suppression of modes two and three in Table~\ref{tab:divergence}. The concentration gains likewise support the analytical capture trend, subject to the mode-one limitation above. These comparisons provide a qualitative cross-check of the launcher and analytical model that informs, rather than replaces, the parametric system analysis.

In summary, the HFSS full-wave simulations confirm three key aspects: (i) the coaxial horn and UCA can be integrated on a shared aperture to produce a far-field pattern that differs measurably from the UCA-only case; (ii) the OAM vortex phase topology is preserved in the presence of the horn; and (iii) the divergence angle is reduced relative to the UCA-only baseline across the three simulated OAM modes, qualitatively supporting the divergence suppression mechanism of the analytical model.


\section{Conclusion}
We have proposed and systematically analyzed a toroidal-OAM hybrid beam that uses far-field features of toroidal electromagnetic pulses (angular confinement and longitudinal electric field) produced by a coaxial horn to suppress the divergence and fill the central null of conventional OAM beams. The concept has been validated through both analytical modeling and full-wave electromagnetic simulation in ANSYS HFSS, where the hybrid launcher demonstrated measurable divergence reduction and preserved OAM vortex phase topology across multiple modes. A closed-form far-field model was formulated, parameterized by dimensionless hardware-linked quantities. Under the optimized strong-coupling configuration ($\alpha=0.7$, $\kappa=1.0$), the hybrid beam reduces divergence by 12--41\% for OAM modes $\ell = 1$--$4$, with modes $\ell \geq 5$ achieving complete divergence suppression (a center-peaked pattern whose intensity peak lies on the axis; the divergence metric tracks the peak position). The hybrid beam delivers 6.4--30.9~dB received SNR gain, fixed-efficiency link-budget range equivalent extension from $+108\%$ to $>3405\%$, and up to five orders of magnitude reduction in multi-mode power imbalance (JFI improving from 0.34 to 0.90 at $z=1000\lambda$). The toroidal modifications are azimuthally symmetric and do not mix OAM modes; the intended-mode power fraction at the ring is $S_{\ell} \ge 0.54$ for modes where the OAM ring is well-defined ($\ell \leq 3$). Central null intensity reaches 8--100\% of the peak, and the OAM-MDM capacity at $z=500\lambda$ increases by up to $2.1\times$ ($107\%$) at SNR=20~dB under per-scheme peak normalization. The divergence suppression ratio is independent of carrier frequency and aperture size, governed solely by the dimensionless product $kR$ and the toroidal confinement parameter $\kappa$ (for fixed mode and fixed $\alpha$, $\beta$, i.e., when the aperture is geometrically scaled with the wavelength at fixed $kR$). These findings establish toroidal beam engineering as a practical and scalable paradigm for OAM-based wireless links at millimeter-wave and terahertz frequencies.  Analytical link and capacity results assume the scalar field model and power convention, fixed-DFT reception, and Section~\ref{sec:analysis}'s per-mode power-fraction surrogate; a unified aperture-integrated receive model and a fully energy-consistent absolute link model remain future work.


\bibliographystyle{IEEEtran}
\bibliography{References}

\end{document}